\documentclass[twocolumn]{aastex631}

\usepackage{graphicx}
\usepackage{subfigure}
\usepackage{paralist}
\usepackage{amssymb}
\usepackage{bm}
\usepackage{bbding}
\usepackage[fleqn]{amsmath}
\usepackage{multirow}

\shorttitle{Tidal coplanarization}
\shortauthors{Lei}

\begin{document}

\title{Tidal coplanarization of circumbinary planetary systems through stellar Cassini states}

\correspondingauthor{Hanlun Lei}
\email{leihl@nju.edu.cn}

\author{Hanlun Lei}
\affiliation{School of Astronomy and Space Science, Nanjing University, Nanjing 210023, China}
\affiliation{Key Laboratory of Modern Astronomy and Astrophysics in Ministry of Education, Nanjing University, Nanjing 210023, China}



\begin{abstract}
Circumbinary planets (CBPs) currently identified are in nearly coplanar configurations relative to their host binaries, yet the dynamical origin of this preference remains unclear. We investigate this question by simulating the secular spin–orbit evolution of CBP systems with tidal decay. A representative case shows that the system evolves through three stages—coplanarization, spin–orbit synchronization, and spin–orbit alignment—through the angular momentum exchange between stellar spin and orbital motion. The evolution of mutual inclination is strongly coupled to stellar obliquity. Phase-space analysis and examination of stellar Cassini states reveal that arbitrary initial inclinations are gradually damped to coplanarity by tides, while stellar obliquity is adiabatically captured into Cassini states with diminishing oscillation amplitudes. This study provides a self-consistent analytical and numerical framework for determining stellar Cassini states and understanding coupled spin–orbit evolution in CBP systems. It shows that tidal dissipation, combined with adiabatic capture into Cassini states, drives the observed dynamical behavior.
\end{abstract}



\section{Introduction}
\label{Sect1}  

Stellar binaries and higher-order multiple systems are ubiquitous in our Galaxy, with the fraction reaching up to 70\% for massive stars \citep[e.g.,][]{tokovinin1997msc}. Hundreds of exoplanets have been observed in such systems, with roughly 80\% residing in binaries \citep[e.g.,][]{martin2018populations}. In binary systems, planetary orbits are typically classified as either circumstellar (S-type), around a single star, or circumbinary (P-type), encompassing both stars \citep[e.g.,][]{cuntz2014s, cuntz2015s, wang2019s}. Planets in P-type orbits, known as circumbinary planets (CBPs), offer unique laboratories for studying planet formation and dynamical evolution in multi-star environments. The dynamics of circumbinary planetary systems was thoroughly investigated in \citet{lei2024Dynamical}, who showed that the quadrupole-order (nodal) resonance shapes the V-shaped region, with high-order and secondary resonances dominating the structures inside and outside it, while the secondary 1:1 resonance is responsible for breaking the symmetry of dynamical structures in the polar region.

To date, over 30 CBPs have been confirmed through various detection techniques\footnote{See \url{ http://exoplanet.eu/} and \url{https://exoplanet.eu/planets_binary_circum/}.}, including planetary transits \citep[e.g.,][]{doyle2011kepler}, direct imaging \citep[e.g.,][]{burgasser2010clouds, kuzuhara2011widest}, eclipse timing variations \citep[e.g.,][]{baran2015detection, borkovits2016comprehensive, getley2017evidence, goldberg20235}, microlensing \citep[e.g.,][]{bennett2016first, kuang2022ogle}, and radial velocity monitoring \citep[e.g.,][]{baycroft2025bebop}. Among these, approximately a dozen are discovered by the Kepler \citep[e.g.,][]{orosz2012kepler, orosz2012neptune, orosz2019discovery, welsh2012transiting, welsh2015kepler, schwamb2013planet, kostov2014kepler, kostov2016kepler, socia2020kepler} and TESS \citep[e.g.,][]{kostov2020toi, kostov2020multiple, kostov2021tic} space telescopes. Many circumbinary systems are observed within 1 AU, where tidal dissipation can gradually influence binary dynamics \citep[e.g.,][]{macdonald1964tidal,hut1981tidal,ogilvie2014tidal,wang2019orbital}. In particular, it can slowly modify the spins and orbital elements of the inner binary pair, and thus significantly alter the final configuration of systems \citep[e.g.,][]{correia2011tidal,correia2012pumping,correia2013tidal,correia2015spin,shevchenko2018tidal,farhat2025capture,liu2026planetary}.

A notable feature of CBPs currently detected is the apparent absence of transiting planets around compact binaries with orbital periods shorter than 7 days, a phenomenon known as ``circumbinary planet desert” \citep[e.g.,][]{orosz2012kepler, liu2026planetary}. Several mechanisms have been proposed to explain this desert, including the von Zeipel–Lidov–Kozai (ZLK) effect \citep[e.g.,][]{kozai1962secular, lidov1962evolution, fabrycky2007shrinking, martin2015no, naoz2016eccentric, moe2018dynamical,lei2022systematic}, evection resonances \citep[e.g.,][]{touma1998resonances, xu2016disruption}, magnetic braking \citep[e.g.,][]{verbunt1981magnetic}, tidal synchronization \citep[e.g.,][]{fleming2018lack}, observational concealment \citep[e.g.,][]{mogan2025concealing}, apsidal precession resonance \citep[e.g.,][]{farhat2025capture}, and dynamical instability induced by planet–planet scattering \citep[e.g.,][]{liu2026planetary}. Another striking characteristic of the currently known CBPs is that they orbit their host binaries on nearly coplanar planes, with many located close to the dynamical stability limit \citep[e.g.,][]{armstrong2014abundance, windemuth2019modelling, georgakarakos2024empirical}. While this preference for coplanarity may largely result from observational biases, it does not necessarily reflect the intrinsic properties of the underlying CBP population \citep[e.g.,][]{czekala2019degree, martin2019binary, childs2021formation, lei2024Dynamical}.

Recent observations have revealed misaligned gas or debris disks around several stellar binaries, including 99 Herculis \citep[e.g.,][]{kennedy201299}, IRS 43 \citep[e.g.,][]{brinch2016misaligned}, GG Tau \citep[e.g.,][]{cazzoletti2017testing}, HD 142527 \citep[e.g.,][]{verhoeff2011complex}, HD 98800B \citep[e.g.,][]{kennedy201299}, and AC Her \citep[e.g.,][]{martin2023ac}, indicating that planet formation in misaligned configurations is possible. From a dynamical perspective, polar orbits in P-type configurations are stable across a broad range of binary parameters \citep[e.g.,][]{cuello2019planet, chen2020polar, lei2024Dynamical}. Moreover, $N$-body simulations suggest that terrestrial planets are more likely to form around eccentric binaries in polar orbits than in coplanar configurations \citep[e.g.,][]{childs2021formation}. These findings imply that misaligned (even polar) orbits can provide favorable dynamical environments for planet formation.

From the perspective of planet formation, the possibility of misaligned orbits cannot be excluded. This raises a natural question: is there a dynamical mechanism underlying the apparent preference for coplanar configurations among observed CBPs? Within the framework of three-body circumbinary systems with tidal effects, \citet{correia2016secular} showed that circumbinary planets with arbitrary orbital inclinations can evolve toward coplanar configurations through a secular resonance between the precession of the inner orbit and the precession of the stellar spin. The goal of this work is to systematically investigate the mechanism of tidal coplanarization. To this end, we study the secular and tidal evolution of circumbinary systems within a non-restricted hierarchical three-body framework, incorporating octupole-order third-body perturbations, general relativistic corrections, rotational and tidal bulges, and tidal dissipation. Our numerical simulations show that the system first evolves toward coplanarization, then toward spin–orbit synchronization, and finally achieves spin–orbit alignment through the exchange of orbital and rotational angular momentum. The evolution of the mutual inclination is strongly coupled to that of the stellar obliquity. To understand this behavior, we analyze the phase-space structure and stellar Cassini states using Poincar\'e sections and perturbative methods. Our results indicate that arbitrary initial inclinations can be damped to coplanar configurations as a consequence of tidal dissipation, while stellar obliquity is captured adiabatically into Cassini states.

The remainder of this work is organized as follows. In Section \ref{Sect2}, the dynamical model with tides used in this work is briefly introduced and a representative example is presented. Under a conservative approximation, the orbital and spin Hamiltonian models are separately formulated in Section \ref{Sect3}. Dynamical structures and Cassini states are numerically explored in Section \ref{Sect4} and analytically studied in Section \ref{Sect5}. In Section \ref{Sect6}, the results in conservative systems are applied to understand the coupled evolution of mutual inclination and stellar obliquity in the presence of tidal effects. At last, the conclusions are summarized in Section \ref{Sect7}.   

\begin{table*}
\centering
\caption{Physical parameters of the considered circumbinary planetary system, which is similar to the Kepler-34 system \citep[e.g.,][]{correia2016secular}.}
\begin{tabular*}{\textwidth}{@{\extracolsep{\fill}}lccc@{\extracolsep{\fill}}}
\hline
{Parameters}&{Star A ($i=0$)}&{Star B ($i=1$)}&{Planet ($i=2$)}\\
\hline
Mass $m_{i}$ ($m_{\sun}$)&1.0&0.2&0.001\\
Spin period $P_{i}$ (day)&10&1.0&0.5\\
Physical radius $R_{i}$ ($\times 10^{3}\,{\rm km}$)&695&150&70\\
The time lag $\Delta t_{i}$ (s)&0.1&0.1&100\\
The square of radius of gyration $r_{g,i}^{2}$&0.08&0.08&0.25\\
The second Love number $k_{2,i}$&0.028&0.028&0.5\\
\hline
\end{tabular*}
\label{Tab1}
\end{table*}

\begin{figure}
\centering
\includegraphics[width=0.8\columnwidth]{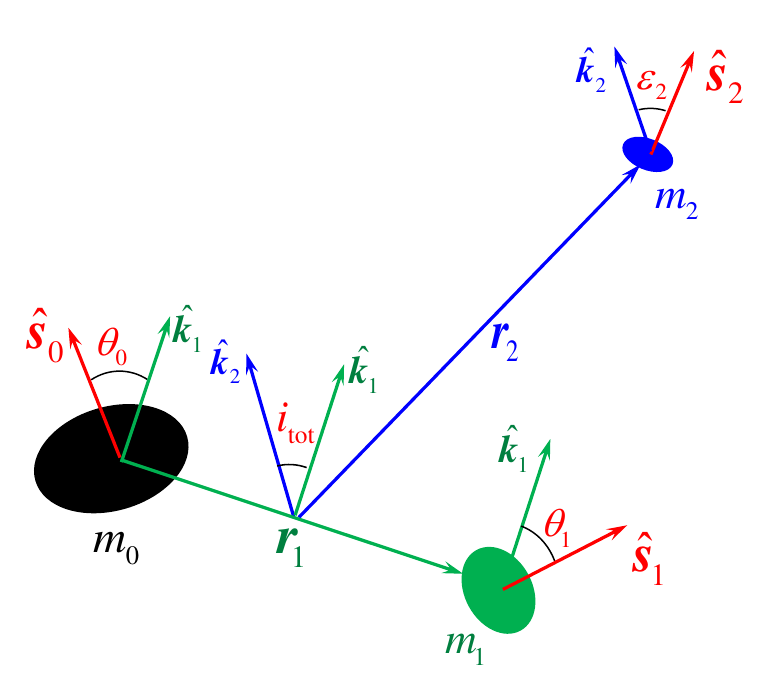}
\caption{Geometric configuration of a circumbinary planetary system, which is composed of an inner stellar binary with mass of $m_0$ and $m_1$ and a faraway planet with mass of $m_2$. The motion of the three objects are described by Jacobi coordinates, where ${\bm r}_1$ is the position of $m_1$ relative to $m_0$ (corresponding to inner binary orbit), and ${\bm r}_2$ is the position of $m_2$ relative to the barycenter of $m_0$ and $m_1$ (corresponding to outer binary orbit). All the bodies are considered as oblate ellipsoids, with spin axes ${\hat {\bm s}_0}$, ${\hat {\bm s}_1}$ and ${\hat {\bm s}_2}$ along their maximum inertial axes. The unitary vectors of orbital angular momentum are denoted by ${\hat {\bm k}_1}$ for the inner orbit and ${\hat {\bm k}_2}$ for the outer orbit. The mutual inclination between the inner and outer orbits is denoted by $i_{\rm tot}$, the obliquity of the stellar binary $m_0$ and $m_1$ relative to the inner orbit are denoted by $\theta_0$ and $\theta_1$, and the obliquity of the planet $m_2$ relative to the outer orbit is denoted by $\varepsilon_2$.}
\label{Fig1}
\end{figure}

\section{Secular and tidal evolutions}
\label{Sect2}

\subsection{Dynamical model}
\label{Sect2-1}

The purpose of this work is to study secular spin-orbit dynamics of circumbinary planetary systems. To this end, we consider a hierarchical three-body problem, which is composed of an inner stellar binary with masses of $m_0$ and $m_1$, together with a faraway planet with mass of $m_2$. Without loss of generality, it is assumed that $m_0 \ge m_1 \gg m_2$. For convenience, we refer to the body $m_0$ as the primary star and the one $m_1$ as the secondary star. From a general point of view, all the bodies are considered as oblate ellipsoids. Under the gyroscopic approximation, oblate ellipsoids are assumed to rotate about their maximum inertial axes along the direction of ${\hat{{\bm s}_i}}$ with rotation velocity of $\Omega_{{\rm rot},i}$, where the subscripts $i=0,1,2$ are used to specify the variables of $m_0$, $m_1$ and $m_2$, respectively. The gravity field coefficients of the rotating oblate bodies are given by
\begin{equation*}
J_{2,i} = k_{2,i} \frac{\Omega_{{\rm rot},i}^2 R_i^3}{3{\cal G} m_i},\quad i=0,1,2
\end{equation*}
where ${\cal G}$ is the universal gravitational constant, $R_i$ is the radius of the $i$-th body, $k_{2,i}$ is the associated second Love number for the potential.

To formulate the dynamical model, we adopt the Jacobi coordinates, where ${\bm r}_1$ is the position vector of $m_1$ relative to $m_0$, ${\bm r}_2$ is the position vector of $m_2$ relative to the barycenter of the inner binary. We assume that the system is hierarchical, thus it holds $r_1 \ll r_2$. The unitary vectors of orbital angular momentum are denoted by ${\hat{\bm k}_1}$ for the inner orbit and ${\hat{\bm k}_2}$ for the outer orbit. Accordingly, the mutual inclination between the inner and outer orbits is specified by
\begin{equation*}
\cos{i_{\rm tot}} = {\hat{\bm k}_1} \cdot {\hat{\bm k}_2}.
\end{equation*}
The relative angle between the rotational and orbital angular momentum vectors is referred to as obliquity, which is denoted by $\theta_0$ for the primary star, $\theta_1$ for the secondary star and $\varepsilon_2$ for the planet. According to the geometry, it holds
\begin{equation*}
\cos{\theta_0}={\hat {\bm s}_0} \cdot {\hat {\bm k}_1},\quad \cos{\theta_1}={\hat {\bm s}_1} \cdot {\hat {\bm k}_1},\quad \cos{\varepsilon_2}={\hat {\bm s}_2} \cdot {\hat {\bm k}_2}.
\end{equation*}
Please refer to Figure \ref{Fig1} for the relative configuration.

In circumbinary planetary systems, orbital states of the intrested bodies can be characterized by the orbital angular momenta ${\bm G}_{1,2}$ and the Laplace--Runge--Lenz vectors ${\bm e}_{1,2}$ \citep[e.g.,][]{correia2016secular}
\begin{equation}\label{Eq1}
{{\bm G}_i} = {G_i}{\hat {\bm k}_i},\quad {G_i} = {\beta _i}\sqrt {{\mu _i}{a_i}\left( {1 - e_i^2} \right)},\quad i=1,2
\end{equation}
and
\begin{equation}\label{Eq2}
{{\bm e}_i} = \frac{{{{\dot {\bm r}}_i} \times {{\bm G}_i}}}{{{\beta _i}{\mu _i}}} - \frac{{{{\bm r}_i}}}{{{r_i}}},\quad i=1,2
\end{equation}
where $a_i$ is the semimajor axis, $e_i$ is the eccentricity, and $G_i$ is the magnitude of orbital angular momentum. Additionally, $\beta_{1,2}$ are the reduced masses, given by
\begin{equation*}
\beta_1 = \frac{m_0 m_1}{m_0 + m_1},\quad \beta_2 = \frac{m_2 (m_0 + m_1)}{m_0 + m_1 + m_2},
\end{equation*}
and $\mu_{1,2}$ are the gravitational parameters, defined by
\begin{equation*}
\mu_1 = {\cal G} (m_0 + m_1),\quad \mu_2 = {\cal G} (m_0 + m_1 + m_2).
\end{equation*}
On the other hand, spin states of the three oblate bodies are described by rotational angular momenta as follows:
\begin{equation*}
{\bm L}_{i} = C_i \Omega_{{\rm rot},i} {\hat{\bm s}}_i,\quad i=0,1,2
\end{equation*}
where $C_i$ is the principal moment of inertia, defined by
\begin{equation*}
C_i = m_i R_i^2 r_{g,i}^2,\quad i=0,1,2 
\end{equation*}
with $r_{g,i}^2$ as the square of the radius of gyration. Please refer to Table \ref{Tab1} for the physical parameters of the considered circumbinary planetary system.

In non-restricted three-body problems, secular evolution equations in terms of ${\bm G}_{1,2}$, ${\bm e}_{1,2}$ and ${\bm L}_{0,1,2}$ have been developed by \citet{correia2011tidal} and generalized by \citet{correia2016secular}. For convenience, they are repeated in Appendix \ref{Appendix_A}. Of course, we could directly integrate equation (\ref{Eq_A1}) to study secular and tidal evolutions. However, in practice it is usual to perform the transformation between ${\bm G}_{1,2}$, ${\bm e}_{1,2}$ and orbital elements. For convenience, we provide an alternative formulation for evolution equations in terms of orbital elements. 

Considering the fact that the orientation of orbits can be determined by the inclination $i_{1,2}$, longitude of ascending node $\Omega_{1,2}$ as well as argument of pericenter $\omega_{1,2}$, we can obtain the averaged equations of motion in terms of orbital elements as follows: 
\begin{equation}\label{Eq3}
\begin{aligned}
\frac{1}{{2{a_1}}}\frac{{{\rm d}{a_1}}}{{{\rm d}t}} &= \frac{1}{{{G_1}}}\left( {{{\dot {\bm G}}_1} \cdot {{\hat {\bm k}}_1}} \right) + \frac{{{e_1}}}{{1 - e_1^2}}\left( {{{\dot {\bm e}}_1} \cdot {{\hat {\bm e}}_1}} \right),\\
\frac{{{\rm d}{e_1}}}{{{\rm d}t}} &= {{\dot {\bm e}}_1} \cdot {{\hat {\bm e}}_1},\\
\frac{{{\rm d}{i_1}}}{{{\rm d}t}} &= \frac{1}{{{e_1}}}\sin {\omega _1}\left( {{{\dot {\bm e}}_1} \cdot {{\hat {\bm k}}_1}} \right) - \frac{1}{{{G_1}}}\cos {\omega _1}\left( {{{\dot {\bm G}}_1} \cdot {{\hat {\bm q}}_1}} \right),\\
\sin {i_1}\frac{{{\rm d}{\Omega _1}}}{{{\rm d}t}} &=  - \frac{1}{{{e_1}}}\cos {\omega _1}\left( {{{\dot {\bm e}}_1} \cdot {{\hat {\bm k}}_1}} \right) - \frac{1}{{{G_1}}}\sin {\omega _1}\left( {{{\dot {\bm G}}_1} \cdot {{\hat {\bm q}}_1}} \right),\\
\frac{{{\rm d}{\omega _1}}}{{{\rm d}t}} &= \frac{1}{{{e_1}}}\left( {{{\dot {\bm e}}_1} \cdot {{\hat {\bm q}}_1}} \right) - \cos {i_1}\frac{{{\rm d}{\Omega _1}}}{{{\rm d}t}}.
\end{aligned}
\end{equation}
Same evolution equations of orbits can be obtained for the outer binary by replacing the subscript 1 by 2. Note that the expressions of ${\dot {\bm G}}_{1,2}$ and ${\dot {\bm e}}_{1,2}$ can be found in equation (\ref{Eq_A1}), and the unitary vectors ${\hat{\bm k}}_{1,2}$, ${\hat{\bm e}}_{1,2}$ and ${\hat{\bm q}}_{1,2}$ are presented in equation (\ref{Eq_A2}). Thus, the right-hand side of equation (\ref{Eq3}) can be finally expressed as functions of orbital elements. The spin evolution equations remain the same as the ones in equation (\ref{Eq_A1}). The orbital and rotational evolution equations together constitute a complete and self-consistent dynamical model, which can be used to perform numerical investigations on long-term spin-orbit evolutions of circumbinary planetary systems.

\begin{figure*}
\centering
\includegraphics[width=\columnwidth]{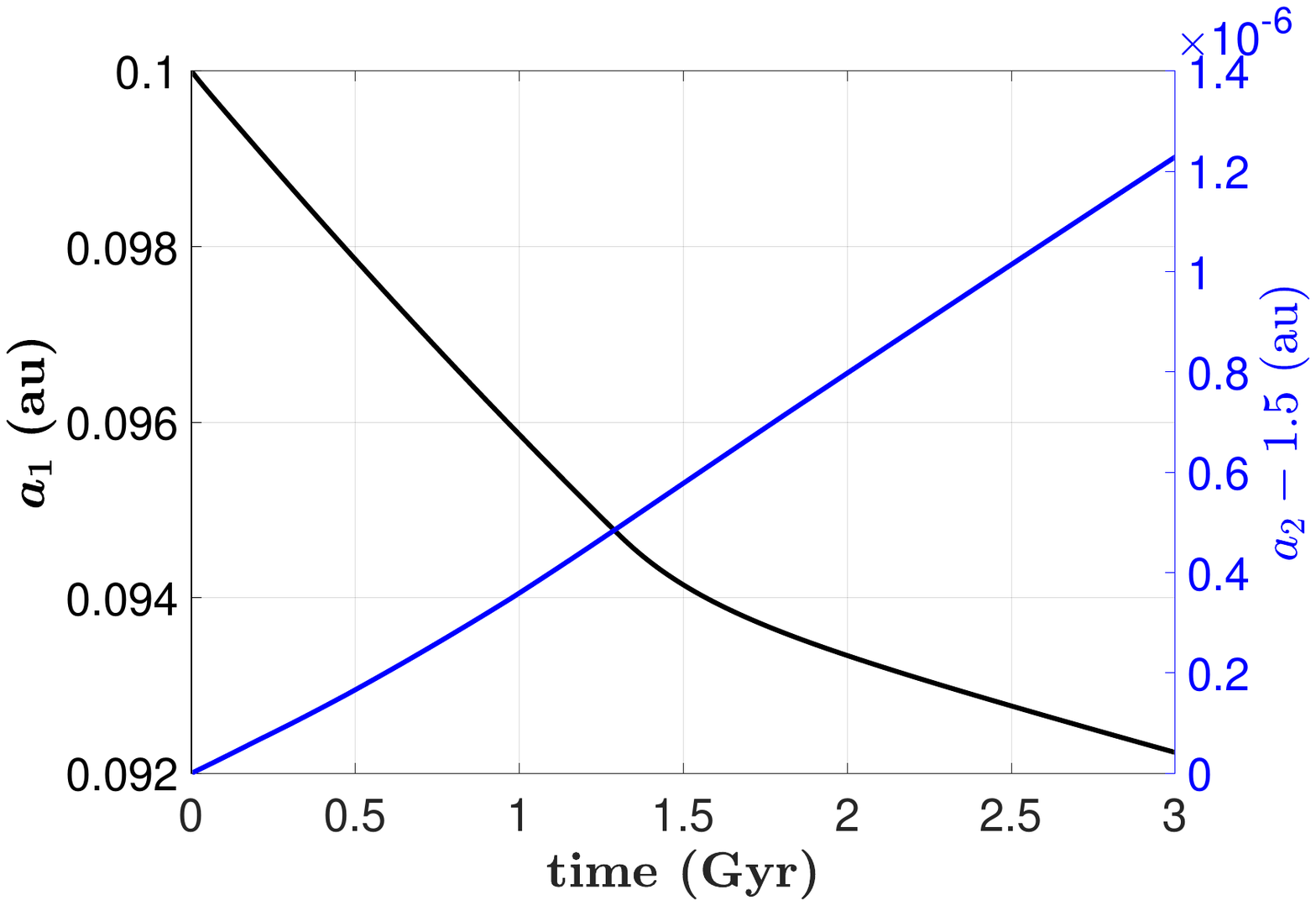}
\includegraphics[width=\columnwidth]{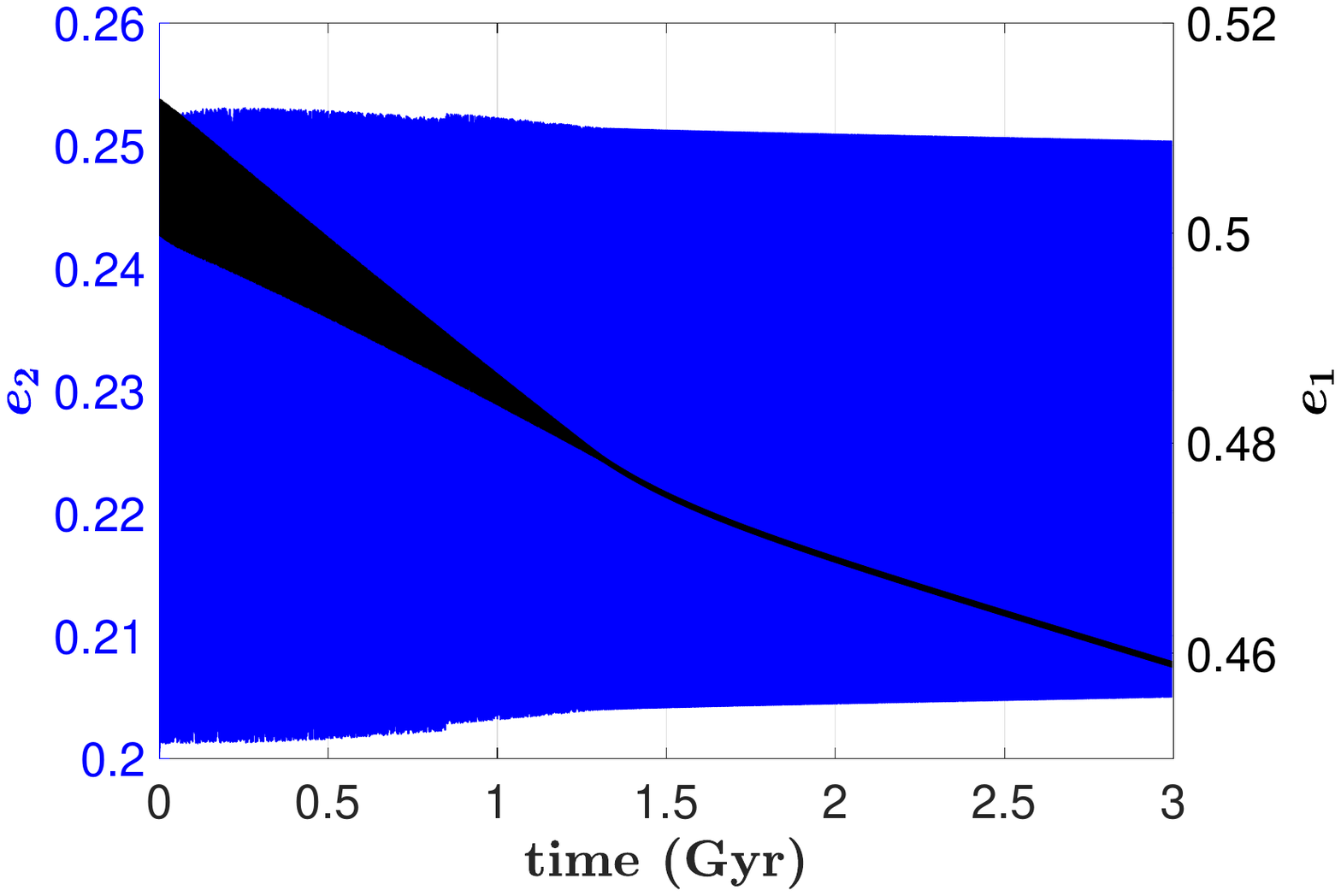}\\
\includegraphics[width=\columnwidth]{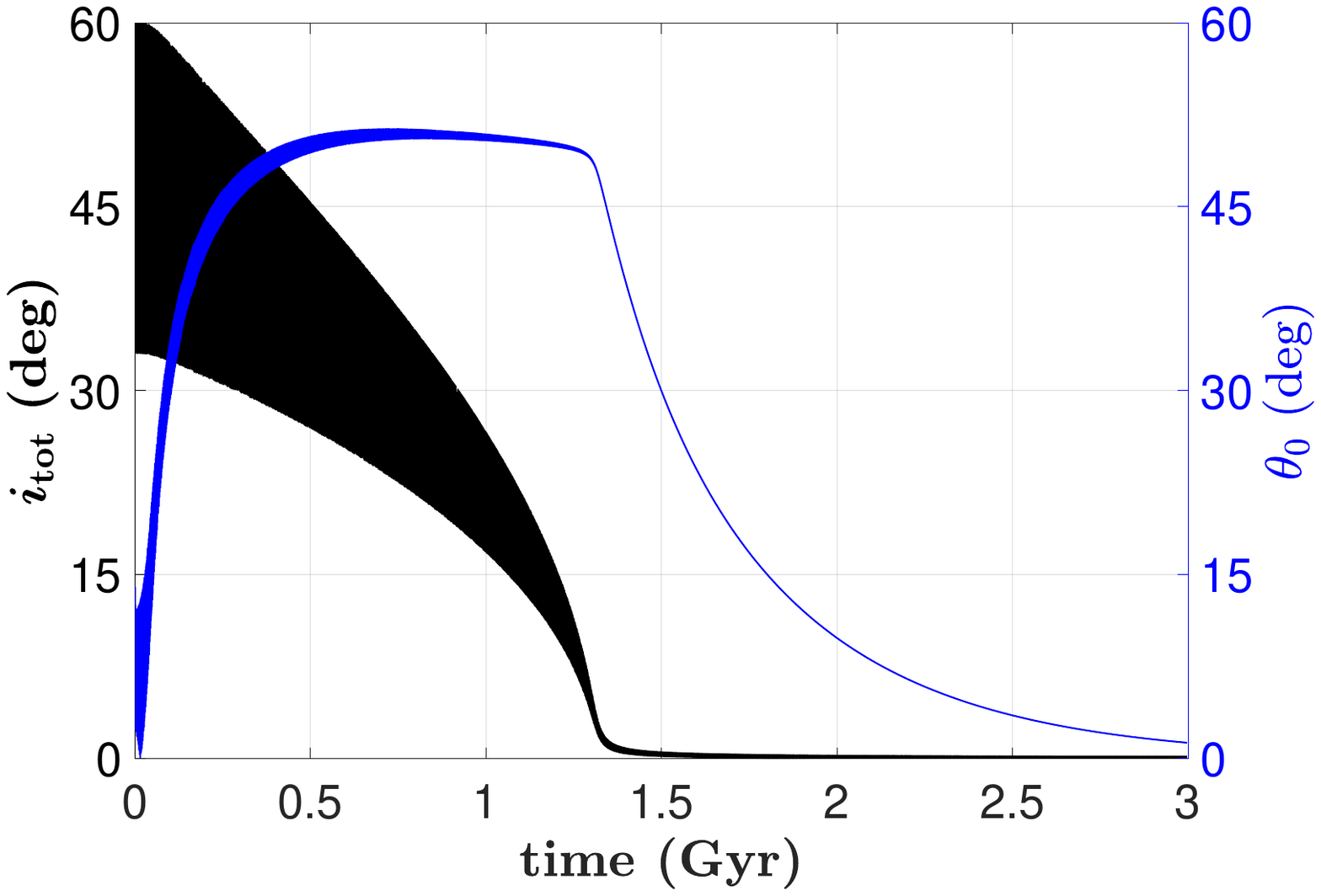}
\includegraphics[width=\columnwidth]{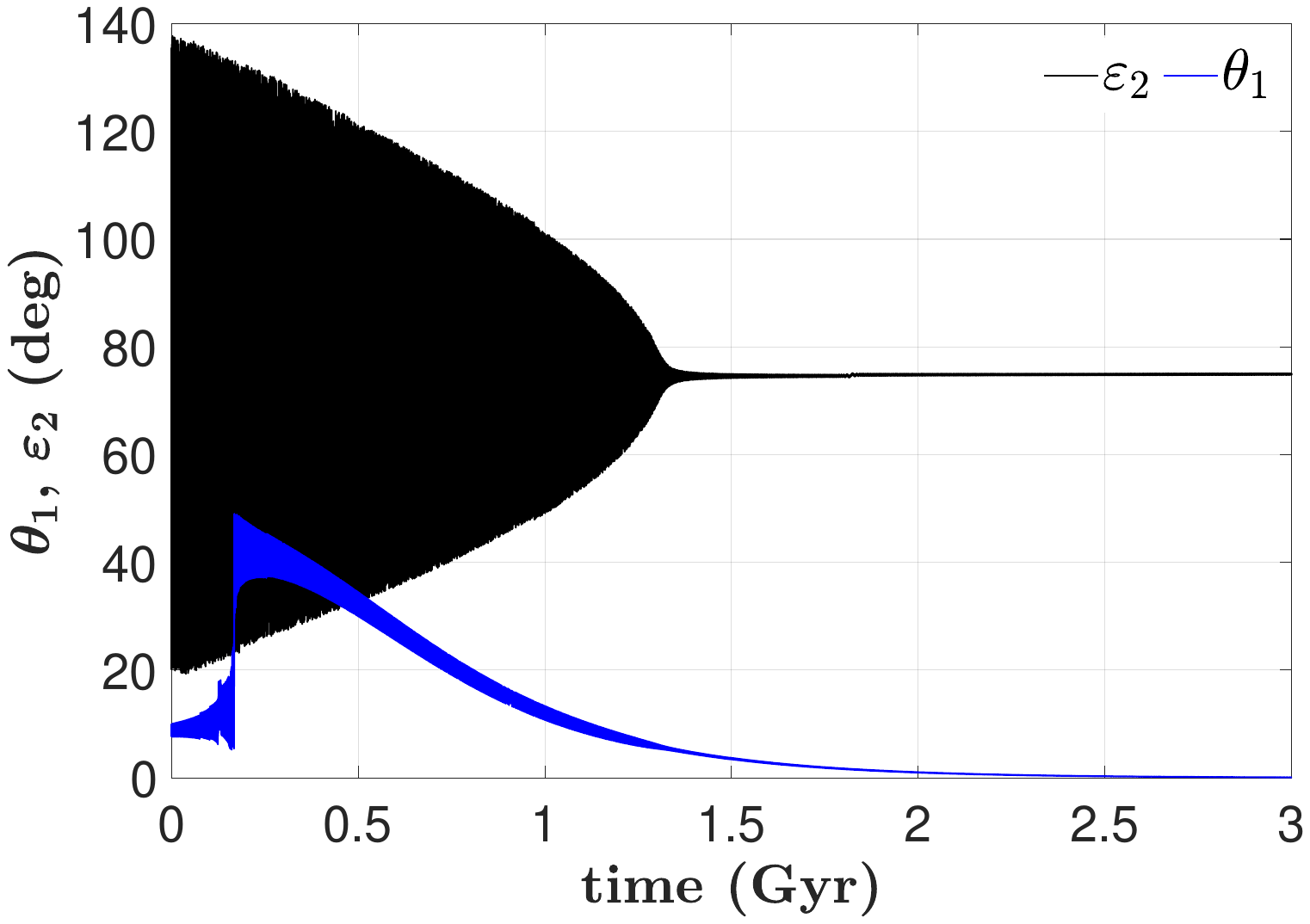}\\
\includegraphics[width=\columnwidth]{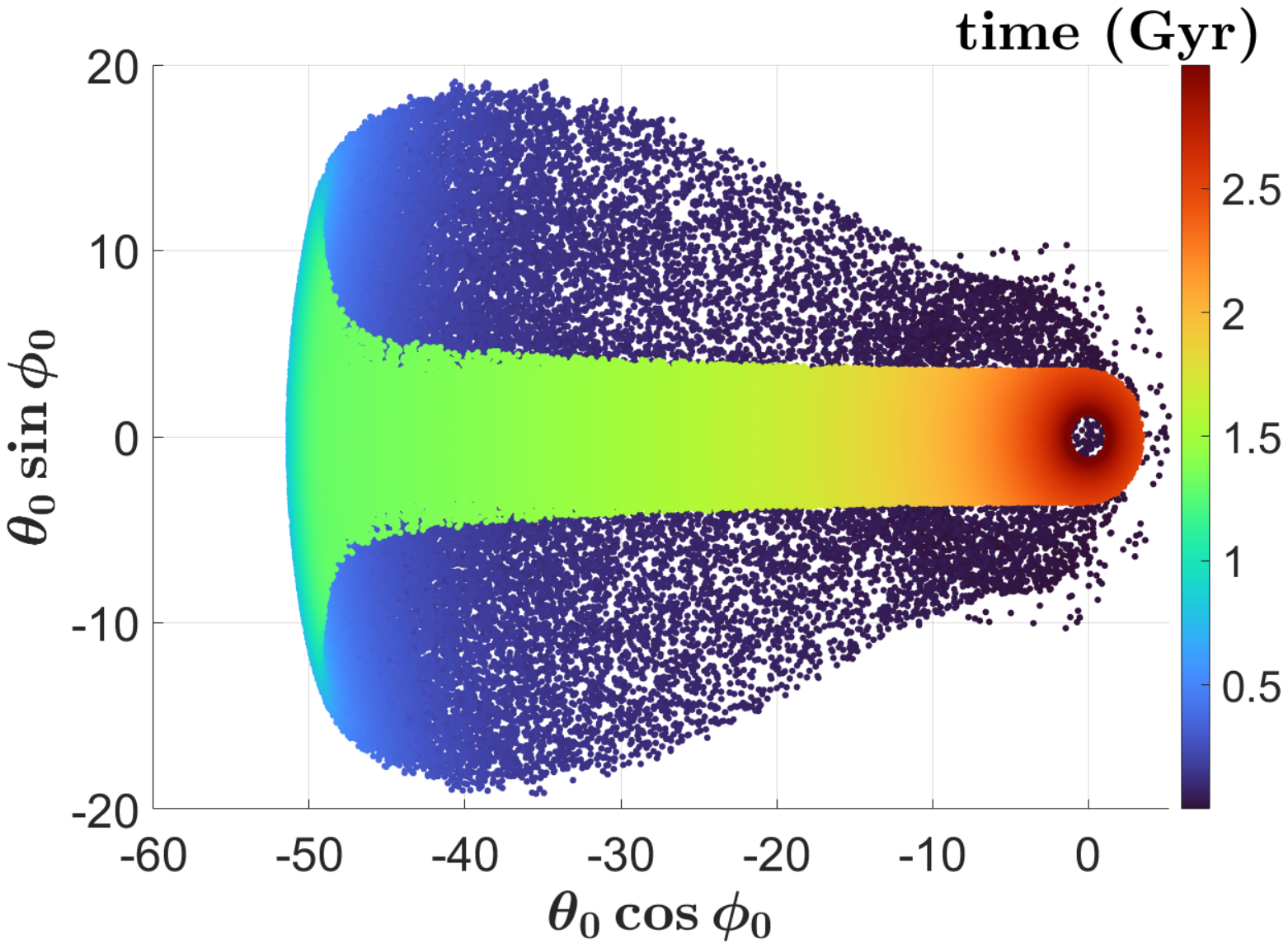}
\includegraphics[width=\columnwidth]{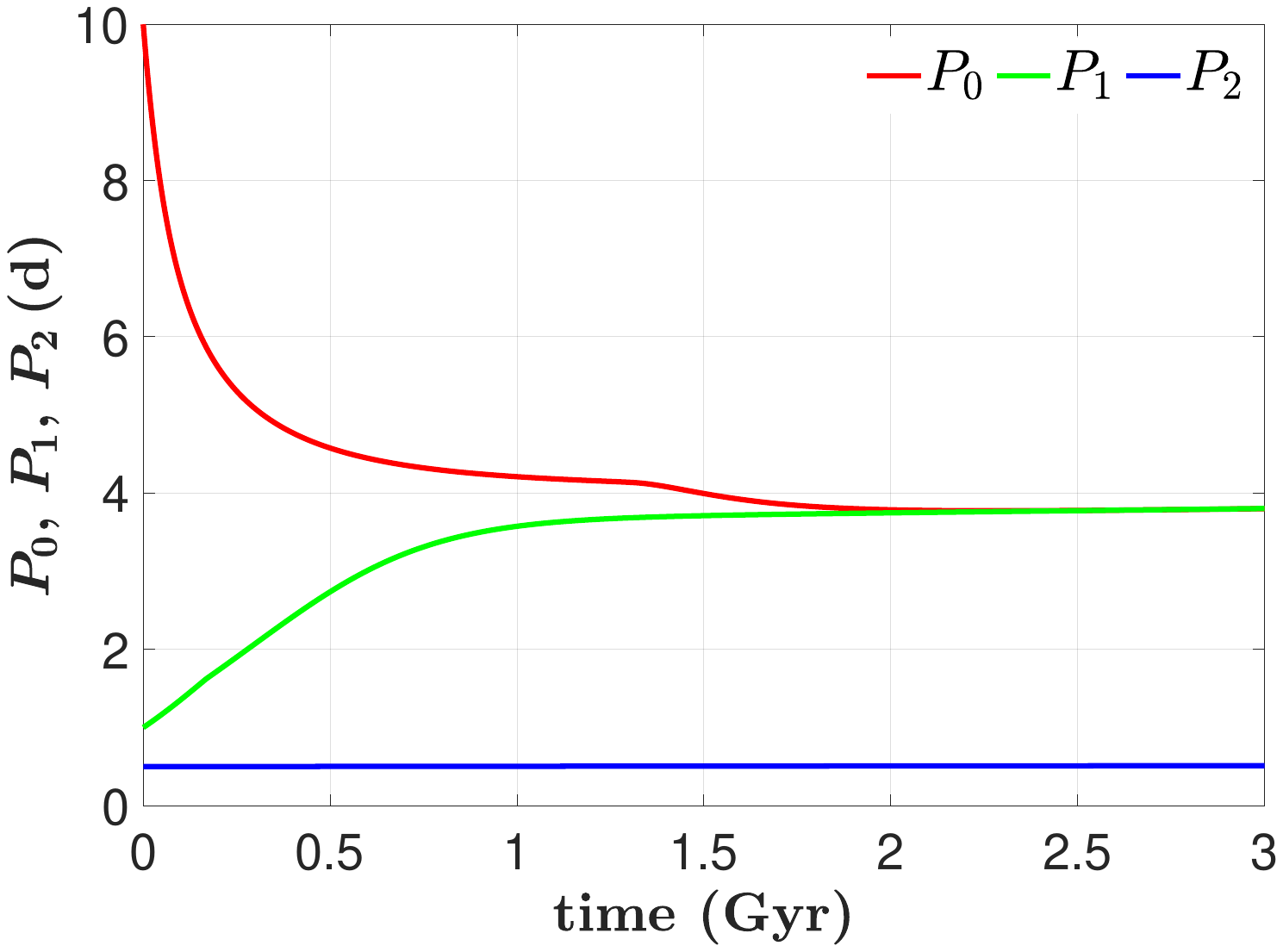}
\caption{Time histories of semimajor axes $a_{1,2}$, eccentricities $e_{1,2}$ and mutual inclination $i_{\rm tot}$ of the inner and outer orbits, the obliquity of the primary star $\theta_0$, the obliquity of the secondary star $\theta_{1}$ and that of the circumbinary planet $\varepsilon_2$, the primary spin projection on the orbital plane $(\theta_0\cos{\phi_0},\theta_0\sin{\phi_0})$, and spin periods $P_{0,1,2}$ for a circumbinary planetary system under the secular octupole-order non-restricted model with consideration of general relativity effect and rotations of all bodies (see Table \ref{Tab1} for the detailed system parameters). The initial inner orbit is characterized by $a_{1,0} = 0.1$ ${\rm au}$ and $e_{1,0}=0.5$, and the outer orbit is specified by $a_{2,0} = 1.5$ ${\rm au}$ and $e_{2,0}=0.2$. The mutual inclination is initially at $i_{\rm tot,0}=60^{\circ}$ and the argument of pericenter of the inner binary is initially assumed at ${\omega_{1,0}}=0$. In this dynamical model, we consider the rotation deformation of all the three objects, with initial obliquity at $\theta_0 = 5^{\circ}$, $\theta_1 = 10^{\circ}$ and $\varepsilon_2 = 20^{\circ}$ and initial phase angles at $\phi_0=\phi_1=\phi_2= 0^{\circ}$. The system parameters and initial conditions are adopted from \citet{correia2016secular}.}
\label{Fig2}
\end{figure*}

\subsection{Numerical examples}
\label{Sect2-2}

To validate the code we use to model spin-orbit evolutions in circumbinary planetary systems, we replicate the results of \citet{correia2016secular}, who studied the secular and tidal evolutions of circumbinary systems, including Kepler-16 and Kepler-34. Here we adopt the same system parameters and initial conditions of their Figure 15. The evolution equations of spins and orbits formulated in the above subsection are numerically integrated over 3.0 Gyr. The results are shown in Figure \ref{Fig2}. Please refer to Table \ref{Tab1} for the setting of system parameters and the caption of Figure \ref{Fig2} for the setting of initial conditions. We can see that the evolutions of $i_{\rm tot}$ and $\theta_0$ are in perfect agreement with Figure 15 of \citet{correia2016secular}\footnote{Their Figure 14 cannot reproduce such a coupled evolution of $i_{\rm tot}$ and $\theta_0$ well because their spin Hamiltonian is available for low-inclination configurations.}. Disclosing the dynamical mechanism underlying such a coupled evolution is the focus of the present study. 

From Figure \ref{Fig2}, it is observed that (a) the semimajor axis of the inner binary shrinks by 8\% and that of the outer binary expands slightly, and both of them have no oscillations during the secular and tidal evolution; (b) there are prominent eccentricity oscillations for both the inner and outer binaries, with decreasing amplitude of oscillation; (c) the mutual inclination $i_{\rm tot}$ decreases from $60^{\circ}$ to zero at about 1.3 Gyr (a timescale of coplanarization), and during this period the obliquity of the primary star increases from the initial $5^{\circ}$ up to $\sim$$50^{\circ}$; (d) after about 1.3 Gyr, the orbital configurations becomes coplanar and the obliquity of the primary star begin to decrease due to the tidal effects, and finally it approaches zero at about 3.0 Gyr; (e) the obliquity of the secondary star $\theta_1$ exhibits a sudden increase at the first 0.2 Gyr up to about $40^{\circ}$ because of a secular spin–orbit resonance and then it gradually decreases to zero at about 2.0 Gyr; (f) there is a significant oscillation for the planetary obliquity $\varepsilon_2$, which switches between prograde and retrograde configurations back and forth, but the amplitude of oscillation decreases with time and it finally converges to a steady state of $\varepsilon_2=78^{\circ}$ at about 1.3 Gyr (from this moment of time the orbital configurations become coplanar); (g) the spin projection of the primary star to the inner orbit as a function of time shows that the spin axis of the primary star is always captured in Cassini states during the entire period of evolution, which may be the dynamical mechanism causing the coupled evolution of $i_{\rm tot}$ and $\theta_0$; (h) the variations of spin periods show that the primary star accelerates rotation and the secondary star decelerates rotation, and both of them arrive at the synchronous spin-orbit state at about 1.8 Gyr; and (i) in the entire duration, the spin period of the planet remains unchanged, meaning that the exchanges between rotational and orbital angular momentum of the outer binary is negligible, which is due to the weak tidal effect in a faraway configuration.    

For the case considered in Figure \ref{Fig2}, we can see that the timescale of inclination damping is about 1.3 Gyr, the timescale of spin-orbit synchronization of the inner binary is about 1.8 Gyr, and the timescale of spin-orbit alignment is about 3.0 Gyr for the primary star and about 2.0 Gyr for the secondary star. It shows that the circumbinary planetary system first evolves to a coplanar state, and then comes to the synchronous state, and at last it arrives at a state of spin-orbit alignment for the inner binary. It is further observed that, at the moment when the orbital configurations are coplanar, the evolution of planetary obliquity converges to a steady state, where the planetary spin axis rotates around the total orbital angular momentum vector in a constant rate with a constant obliquity. According to the eccentricity curves of evolution, it can be predicted that the timescale of orbital circularization is much longer than 3.0 Gyr for both the inner and outer binaries, indicating that the secular evolution is an approximately adiabatic process. 

\begin{figure*}
\centering
\includegraphics[width=\columnwidth]{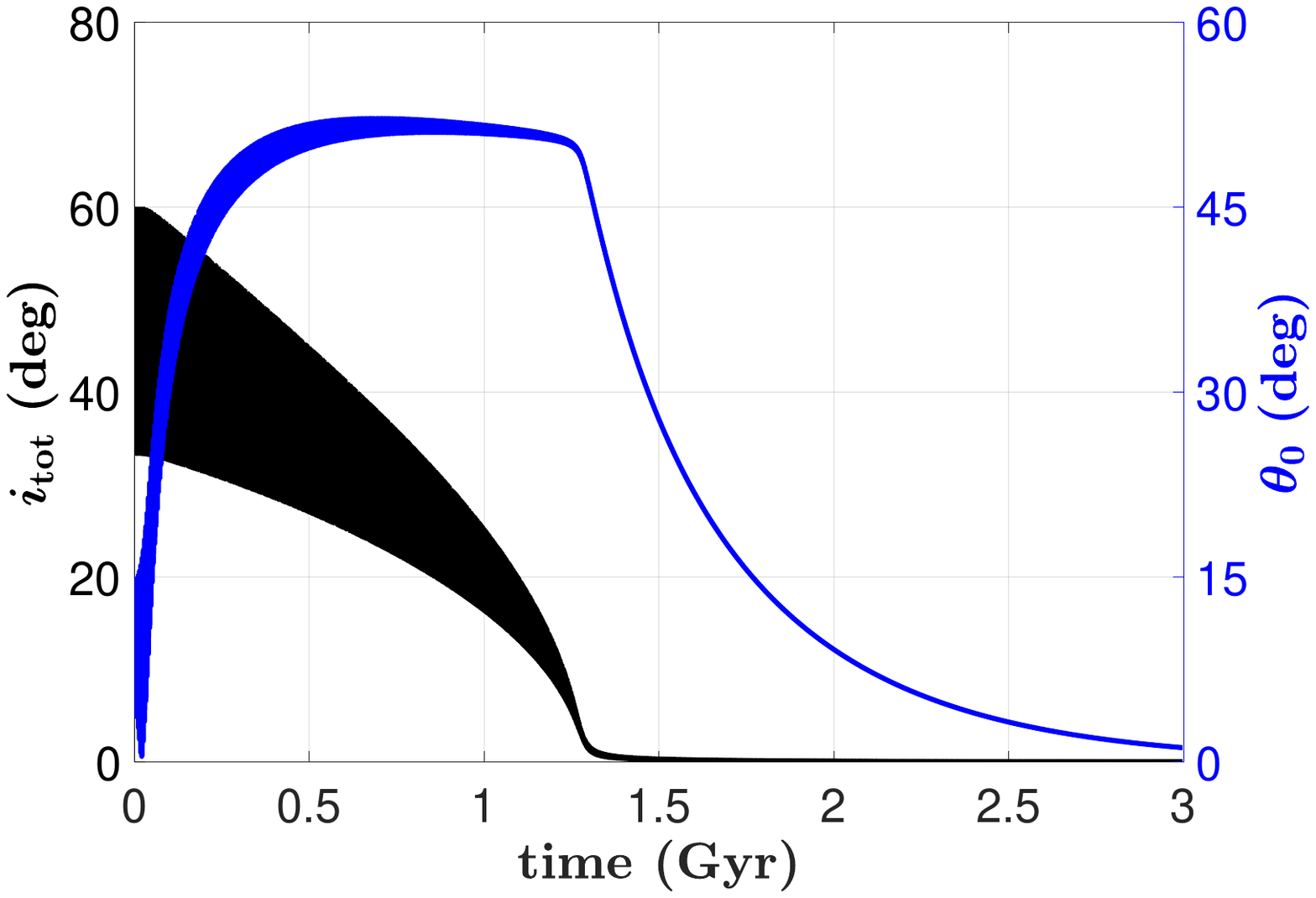}
\includegraphics[width=\columnwidth]{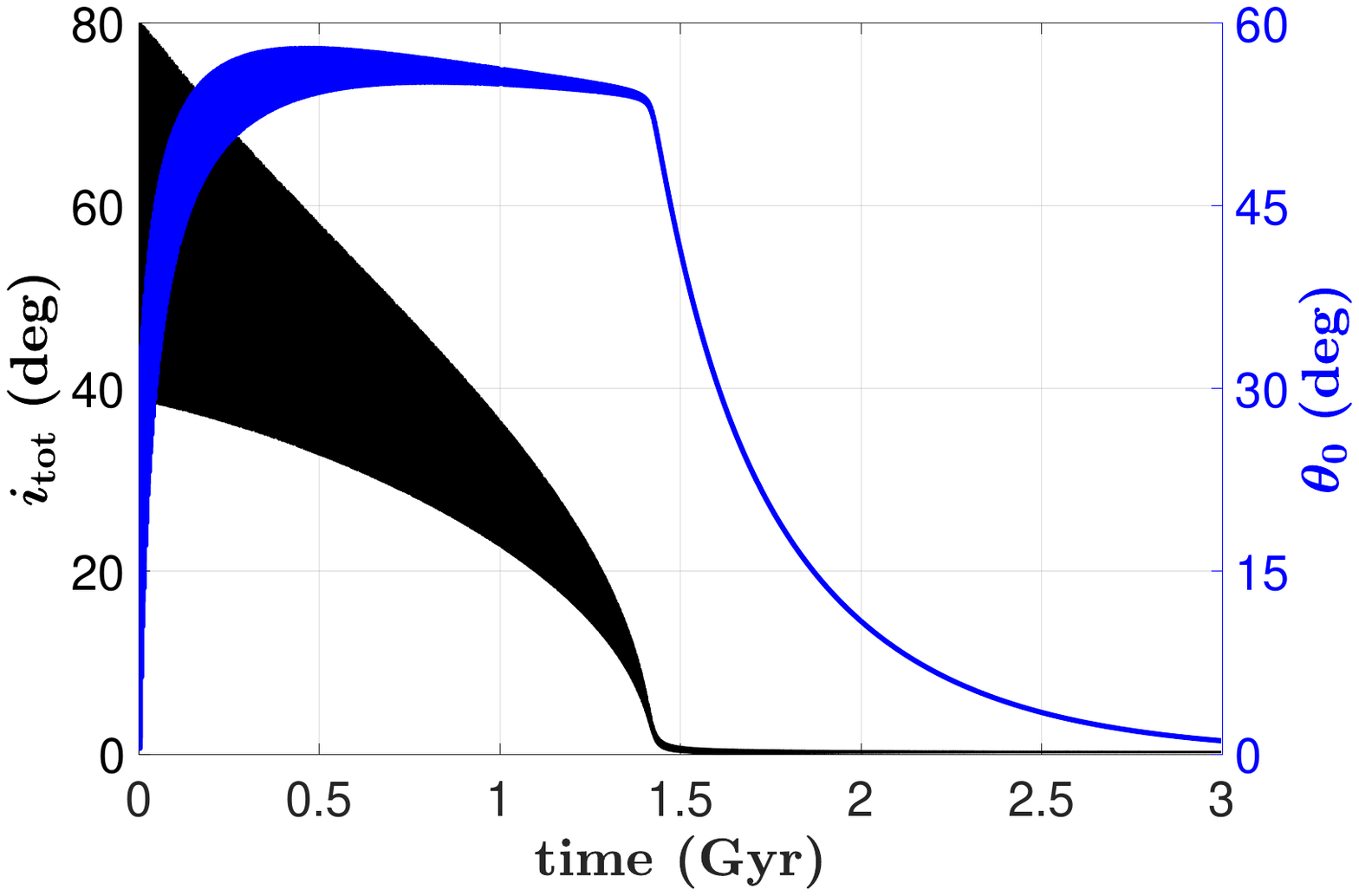}
\caption{Time histories of the mutual inclination $i_{\rm tot}$ and obliquity of the primary star $\theta_0$ for a circumbinary planetary system under the secular quadrupole-order non-restricted problem with general relativity effect and spin of the primary star (see Table \ref{Tab1} for system parameters but assuming $k_{21}=k_{22}=0$). The mutual inclination is initially at $i_{\rm tot,0}=60^{\circ}$ in the left panel and $i_{\rm tot,0}=80^{\circ}$ in the right panel. The remaining initial condition is the same as that adopted in Figure \ref{Fig2}.}
\label{Fig3}
\end{figure*}

It should be mentioned that, in the present study, we are interested in the phenomenon of a coupled evolution of $i_{\rm tot}$ and $\theta_0$. Regarding this point, we may ask a question: Is the coupled evolution robust? To answer this question, we make two extensions. On one hand, we simplify the dynamical model by (a) considering the quadrupole-order approximation of the third-body perturbation, and (b) assuming both the secondary star and the planet as point-mass objects, which is realized by setting their Love numbers as $k_{21}=k_{22}=0$. On the other hand, we change the initial configurations by changing the mutual inclinations. In practice, two representative configurations of $i_{\rm tot,0}=60^{\circ}$ and $i_{\rm tot,0}=80^{\circ}$ are considered in Figure \ref{Fig3}, where the system parameters (except for $k_{21}$ and $k_{22}$) and the initial conditions (except for the initial mutual inclination) are the same as the ones adopted in Figure \ref{Fig2}. Note that we have made more experiments, showing that the coupled evolutions of $i_{\rm tot}$ and $\theta_0$ are qualitatively similar. 

It is observed from Figure \ref{Fig3} that under the simplified model the coupled evolution of $i_{\rm tot}$ and $\theta_0$ remains similar to the one shown in Figure \ref{Fig2}. This means that it is adequate to adopt the simplified model to study the secular spin-orbit evolutions. This is what we do in the following sections. In addition, with a higher magnitude of $i_{\rm tot,0}$, the timescale of inclination damping becomes longer, and the obliquity of the primary star can be excited up to a higher level. However, for both cases, we can see that the timescale of spin-orbit alignment remains almost the same, indicating that the timescale of spin-orbit alignment is insensitive to the initial mutual inclination.

\begin{figure*}
\centering
\includegraphics[width=\columnwidth]{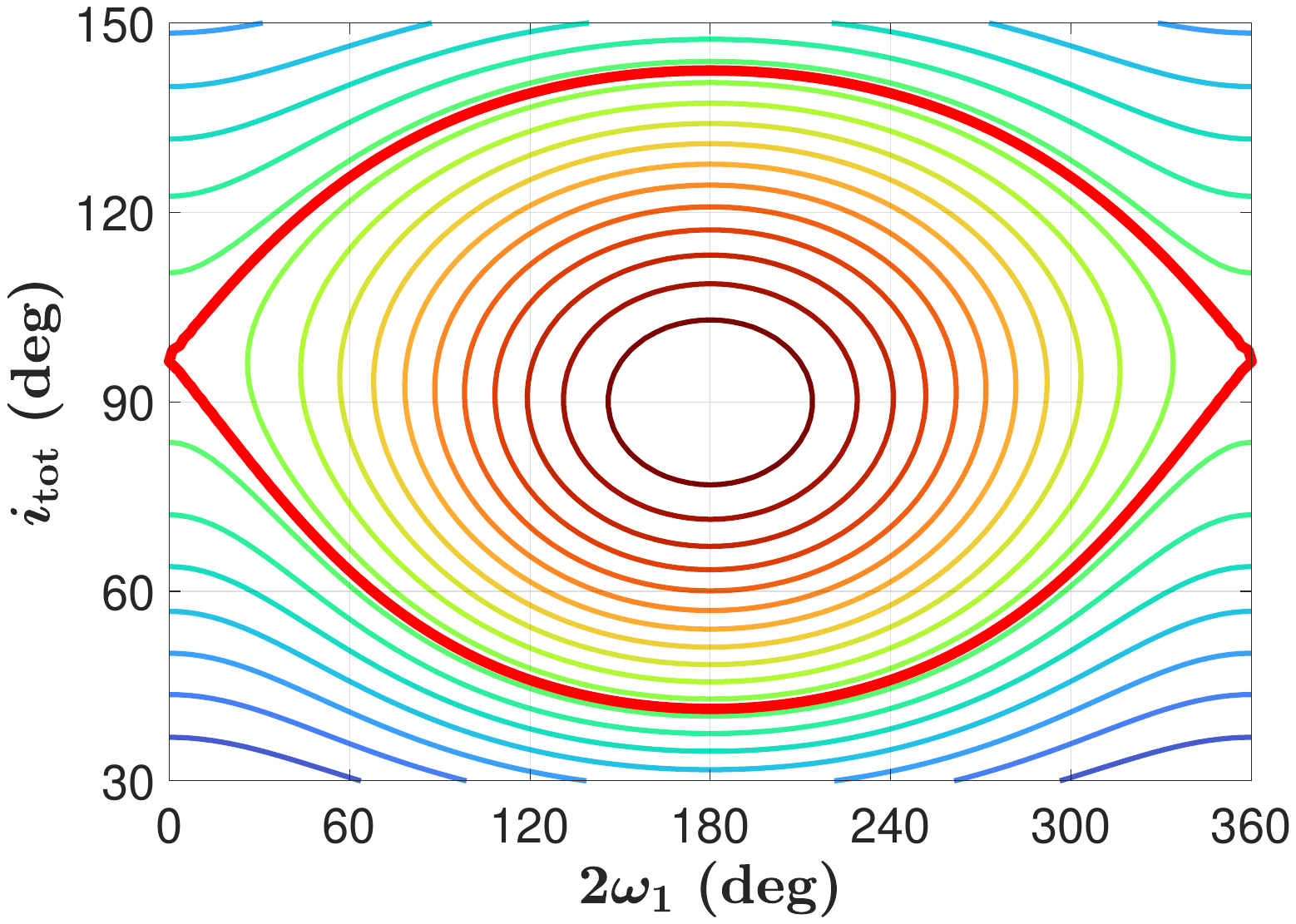}
\includegraphics[width=\columnwidth]{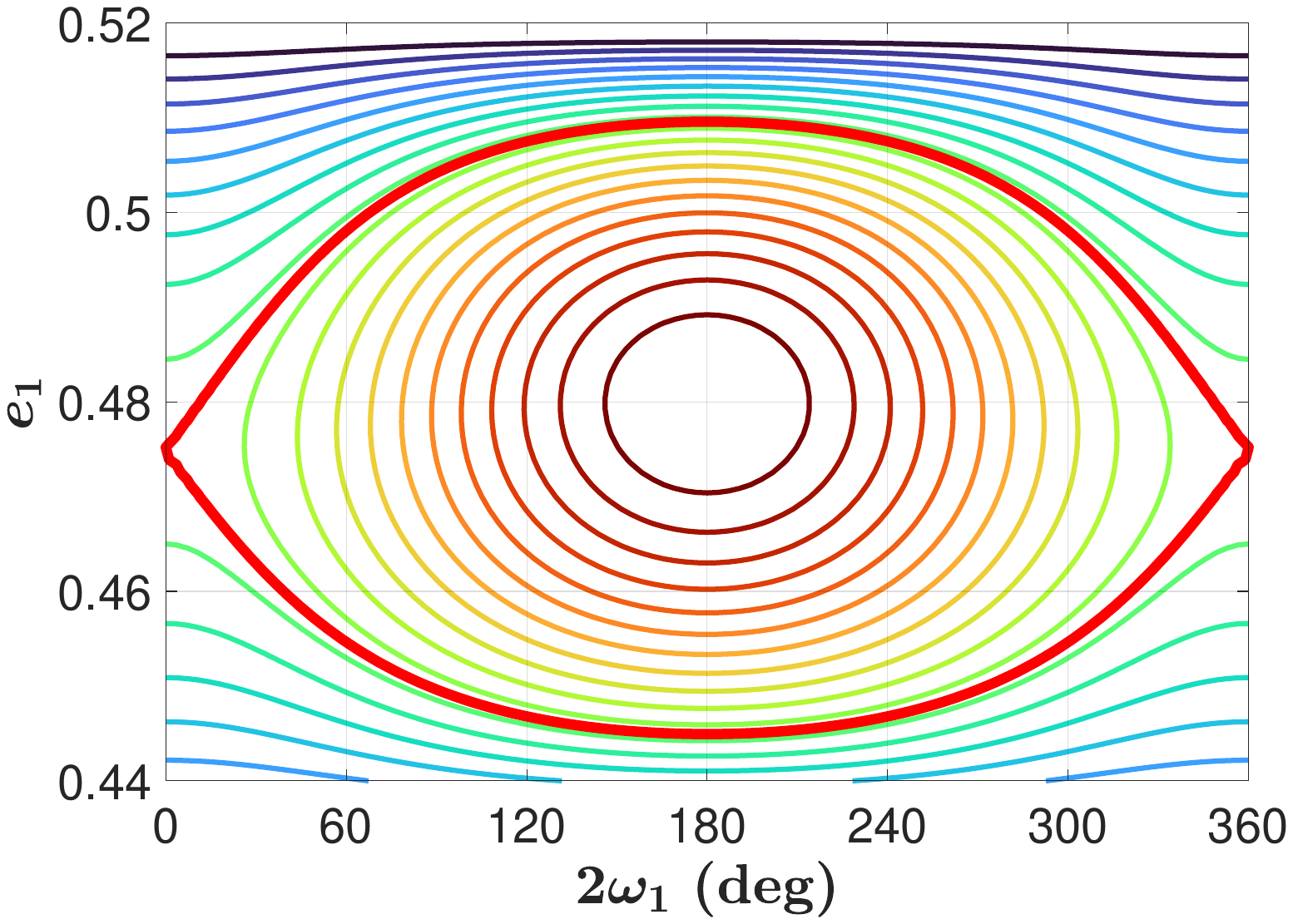}
\caption{Phase portraits (level curves of Hamiltonian) shown in the spaces of $(2\omega_1,i_{\rm tot})$ and $(2\omega_1,e_1)$ for the circumbinary planetary system at the quadrupole-order approximation. The dynamical separatrices passing through saddle points are shown in red lines. The asymmetry structure is due to the inclusion of the general relativity effect of the inner stellar binary.}
\label{Fig4}
\end{figure*}

\section{Hamiltonian formulation of conservative motion}
\label{Sect3}

In the previous section, it is shown that the quadrupole-order approximation of the third-body perturbation and point-mass assumptions of the secondary star and the planet are adequate to produce the coupled evolution of $i_{\rm tot}$ and $\theta_0$. For simplicity, from this section we will adopt such a simplified dynamical model to perform further numerical and analytical explorations. To understand the dynamical behaviors in a dissipative model, we need to investigate the dynamics first in the associated conservative system by turning off tidal effects.

Furthermore, numerical experiments show that the total orbital angular momentum of the inner and outer binaries $G_{\rm tot}$ exhibits a negligible variation during the entire period. This is because the orbital angular momentum $G_{\rm tot}$ is much larger than that of the rotational angular momentum of the primary $L_0$. For example, for the initial conditions of Figure \ref{Fig3}, it holds $L_0/G_{\rm tot} \approx 1.24 \times 10^{-3}$. This observation motivates us to further simplify the dynamical model by decoupling the orbital evolution from the dynamical model, meaning that the influence of the primary spin upon the orbital evolutions of the inner and outer binaries can be ignored. As a result, Hamiltonian models of orbit and spin can be formulated separately.

\subsection{The orbit Hamiltonian}
\label{Sect3-1}

Under the invariant plane coordinate system, the quadrupole-order approximation of Hamiltonian including the general relativistic effect for describing the orbital evolutions of the inner and outer binaries can be written as \citep[e.g.,][]{naoz2013secular}
\begin{equation}\label{Eq4}
\begin{aligned}
{{\cal H}_{\rm quad}} =&  - \frac{1}{3}{\gamma _2}\left[ {\left( {1 + \frac{3}{2}e_1^2} \right)\left( {1 - \frac{3}{2}{{\sin }^2}{i_{\rm tot}}} \right)} \right.\\
&\left. { + \frac{{15}}{4}e_1^2{{\sin }^2}{i_{\rm tot}}\cos 2{\omega _1}} \right] - \frac{{3{\beta _1}\mu _1^2}}{{a_1^2{c^2}\sqrt {1 - e_1^2} }},
\end{aligned}
\end{equation}
where $c$ is the speed of light and the coefficient $\gamma_2$ is given by
\begin{equation*}
{\gamma _2} = \frac{{3{\cal G}{m_2}{\beta _1}a_1^2}}{{4a_2^3{{\left( {1 - e_2^2} \right)}^{3/2}}}}.
\end{equation*}
It is known that the quadrupole-order Hamiltonian determines an integrable dynamical model with two motion integrals, given by
\begin{equation}\label{Eq5}
{G_2} = {\beta _2}\sqrt {{\mu _2}{a_2}\left( {1 - e_2^2} \right)} = {\rm const},
\end{equation}
and
\begin{equation}\label{Eq6}
G_{\rm tot}^2 = G_1^2 + G_2^2 + 2{G_1}{G_2}\cos {i_{\rm tot}} = {\rm const},
\end{equation}
which indicates that the eccentricity of the outer binary $e_2$ remains stationary, and $e_1$ is coupled with $i_{\rm tot}$ to conserve the total orbital angular momentum $G_{\rm tot}$.

Introducing a set of canonical variables $(g_1=\omega_1,G_1)$, we can get the evolution equations of orbits using Hamiltonian canonical relations:
\begin{equation*}
\frac{{{\rm d}{g_1}}}{{{\rm d}t}} = \frac{{\partial {{\cal H}_{\rm{quad}}}}}{{\partial {G_1}}},\quad \frac{{{\rm d}{G_1}}}{{{\rm d}t}} =  - \frac{{\partial {{\cal H}_{\rm{quad}}}}}{{\partial {g_1}}},
\end{equation*}
which leads to the evolution equations of $g_1$ and $G_1$, given by
\begin{equation}\label{Eq7}
\begin{aligned}
\frac{{{\rm d}{g_1}}}{{{\rm d}t}} &= \frac{{{\gamma _2}}}{{2{G_1}}}\left\{ {\left( {1 - e_1^2} \right)\left[ {2 + {{\sin }^2}{i_{\rm tot}}\left( {5\cos 2{g_1} - 3} \right)} \right]} \right.\\
&\left. { - \left( {\frac{{{G_1}}}{{{G_2}}} + \cos {i_{\rm{tot}}}} \right)\cos {i_{\rm{tot}}}\left[ {e_1^2\left( {5\cos 2{g_1} - 3} \right) - 2} \right]} \right\}\\
& + \frac{{3\beta _1^2\mu _1^{5/2}}}{{a_1^{3/2}{c^2}}}\frac{1}{{G_1^2}},\\
\frac{{{\rm d}{G_1}}}{{{\rm d}t}} &=  - \frac{5}{2}{\gamma _2}e_1^2{\sin ^2}{i_{\rm tot}}\sin{2{g_1}}.
\end{aligned}
\end{equation}
It is evident that, given the integral of motion $G_2$ and $G_{\rm tot}$, the evolutions of $g_1$ and $G_1$ are periodic functions of time. Furthermore, it is possible to derive the evolution equations of $i_{1,2}$ and $h_{1,2} (=\Omega_{1,2})$ as follows \citep[e.g.,][]{naoz2013secular}:
\begin{equation}\label{Eq8}
\begin{aligned}
\frac{{{\rm d}{i_1}}}{{{\rm d}t}} &= \frac{5}{2}{\gamma _2}\left( {\frac{1}{{{G_{\rm tot}}}}\frac{1}{{\sin {i_1}}} - \frac{1}{{{G_1}}}\cot {i_1}} \right)e_1^2{\sin ^2}{i_{\rm tot}}\sin 2{g_1},\\
\frac{{{\rm d}{h_1}}}{{{\rm d}t}} &=  - \frac{{{\gamma _2}}}{4}\frac{{\sin \left( {2{i_{\rm tot}}} \right)}}{{{G_1}\sin {i_1}}}\left( {2 + 3e_1^2 - 5e_1^2\cos 2{g_1}} \right),
\end{aligned}
\end{equation}
and 
\begin{equation}\label{Eq9}
\begin{aligned}
\frac{{{\rm d}{i_2}}}{{{\rm d}t}} &=  - \frac{5}{2}{\gamma _2}\frac{{\sin {i_{\rm{tot}}}}}{{{G_2}}}e_1^2\sin 2{g_1},\\
\frac{{{\rm d}{h_2}}}{{{\rm d}t}} &=  - \frac{{{\gamma _2}}}{4}\frac{{\sin \left( {2{i_{\rm tot}}} \right)}}{{{G_1}\sin {i_1}}}\left( {2 + 3e_1^2 - 5e_1^2\cos 2{g_1}} \right).
\end{aligned}
\end{equation}
Similarly, the evolutions of $i_{1,2}$ and $h_{1,2}$ are periodic functions of time. It is noted that the geometric relations always hold $i_{\rm tot} = i_1 + i_2$ and $h_1 - h_2 = \pi$, meaning that $\frac{{{\rm d}{h_1}}}{{{\rm d}t}} = \frac{{{\rm d}{h_2}}}{{{\rm d}t}}$, which is satisfied by equations (\ref{Eq8}) and (\ref{Eq9}).

Due to the conservation of Hamiltonian, the evolution trajectories correspond to level curves of quadrupole-order Hamiltonian (the so-called phase portraits). Figure \ref{Fig4} shows phase portraits in the pseudo phase space of $(2\omega_1,i_{\rm tot})$ and $(2\omega_1,e_1)$. The dynamical separatrices passing through saddle points are marked in red lines, which divide the whole phase space into two regimes, including the libration and circulation regions. It is further observed that the phase-space structure is no longer symmetric with respect to $i_{\rm tot}=90^{\circ}$, which is caused by the general relativity effect. This is in agreement with the result obtained by \citet{correia2016secular}.

\subsection{The spin Hamiltonian}
\label{Sect3-2}

Because our primary goal is to study the evolution of obliquity of the primary star $\theta_0$ relative to the inner orbit, it is convenient to work in the rotating reference frame, where the orbital angular momentum vector of the inner orbit ${\hat{\bm k}}_1$ remains constant\footnote{The spin-orbit dynamics of the secondary star can be studied in the same way.}. In this rotating frame, the Hamiltonian takes the following form \citep[e.g.,][]{kinoshita1993motion}:
\begin{equation}\label{Eq10}
{\cal H}_{\rm spin} =  - \frac{1}{2}{\alpha _0}{\cos ^2}{\theta _0} - {\bm R} \cdot {{\bm L}_0},  
\end{equation}
where the coefficient is
\begin{equation*}
{\alpha _0} = \frac{{3{\cal G}{m_0}{m_1}{J_{2,0}}R_0^2}}{{2a_1^3{{\left( {1 - e_1^2} \right)}^{3/2}}}},
\end{equation*}
and the spin angular momentum vector ${{\bm L}_0}$ and the rotation `matrix' $\bm R$ are given by
\begin{equation*}
{{\bm L}_0} = {L_0}\left[ {\begin{array}{*{20}{c}}
{\sin {\theta _0}\sin {\phi _0}}\\
{ - \sin {\theta _0}\cos {\phi _0}}\\
{\cos {\theta _0}}
\end{array}} \right],\quad {\bm R} = \left[ {\begin{array}{*{20}{c}}
{\frac{{{\rm d}{i_1}}}{{{\rm d}t}}}\\
{\frac{{{\rm d}{h_1}}}{{{\rm d}t}}\sin {i_1}}\\
{\frac{{{\rm d}{h_1}}}{{{\rm d}t}}\cos {i_1}}
\end{array}} \right].
\end{equation*}
As a result, the Hamiltonian for describing the stellar spin dynamics can be organized as
\begin{equation}\label{Eq11}
\begin{aligned}
{\cal H_{\rm spin}} &=  - \frac{1}{2}{\alpha _0}{\cos ^2}{\theta _0} - {L_0}\left[ {\sin {\theta _0}\sin {\phi _0}\frac{{{\rm d}{i_1}}}{{{\rm d}t}}} \right.\\
&\left. { + \left( {\cos {\theta _0}\cos {i_1} - \sin {\theta _0}\cos {\phi _0}\sin {i_1}} \right)\frac{{{\rm d}{h_1}}}{{{\rm d}t}}} \right].
\end{aligned}
\end{equation}
Substituting equation (\ref{Eq8}) into equation (\ref{Eq11}) and normalizing it by $L_0$, we can get the normalized spin Hamiltonian as follows:
\begin{equation}\label{Eq12}
\begin{aligned}
{{\cal H}_{{\rm{spin}}}} &=  - \frac{1}{2}\frac{{\alpha _0}}{L_0} {\cos ^2}{\theta _0} - \frac{5}{2}{\gamma _2}\frac{{e_1^2}}{{\sin {i_1}}}{\sin ^2}{i_{\rm{tot}}} \sin\left({2{g_1}}\right) \\
&\times \left( {\frac{1}{{{G_{\rm{tot}}}}} - \frac{1}{{{G_1}}}\cos {i_1}} \right) \sin {\theta _0}\sin {\phi _0}\\
&+ \frac{1}{4}{\gamma _2}\frac{{\sin \left( {2{i_{\rm tot}}} \right)}}{{{G_1}\sin {i_1}}}\left( {2 + 3e_1^2 - 5e_1^2\cos{2{g_1}}} \right)\\
&\times \left( {\cos {\theta _0}\cos {i_1} - \sin {\theta _0}\cos {\phi _0}\sin {i_1}} \right),
\end{aligned}
\end{equation}
Introducing the set of canonical variables
\begin{equation*}
{x_0} = {\phi _0},\quad {X_0} = \cos {\theta _0},
\end{equation*}
we can rewrite the spin Hamiltonian as
\begin{equation}\label{Eq13}
\begin{aligned}
{{\cal H}_{\rm{spin}}} &=  - \frac{1}{2}\frac{{{\alpha _0}}}{{{L_0}}}X_0^2 - \frac{5}{2}{\gamma _2}\frac{1}{{{G_1}}}e_1^2{\sin ^2}{i_{\rm{tot}}} \sin\left({2{g_1}}\right) \\
&\times \left( {\frac{{{G_1}}}{{{G_{\rm{tot}}}\sin {i_1}}} - \cot {i_1}} \right) \sqrt {1 - X_0^2} \sin {x_0}\\
&+ \frac{1}{4}{\gamma _2}\frac{1}{{{G_1}}}\sin \left( {2{i_{\rm tot}}} \right)\left({2 + 3e_1^2 - 5e_1^2\cos{2{g_1}}} \right)\\
&\times \left( {{X_0}\cot {i_1} - \sqrt {1 - X_0^2} \cos {x_0}} \right).
\end{aligned}
\end{equation}
It is noted that the spin Hamiltonian is closely related to the evolution of inner orbit. However, it is known that the inner orbit variables including $e_1$, $i_1$, $i_{\rm tot}$ and $g_1$ are all periodic functions of time, which are determined by the Hamiltonian ${\cal H}_{\rm quad}$ given by equation (\ref{Eq4}). As a result, the spin Hamiltonian ${\cal H}_{\rm spin}$ given by equation (\ref{Eq13}) is non-autonomous and it determines a 1.5 degree-of-freedom (DOF) dynamical model.

Hamiltonian canonical relation leads to the following spin evolution equations:
\begin{equation*}
\begin{aligned}
{{\dot x}_0} =&  - \frac{{{\alpha _0}}}{{{L_0}}}{X_0} + \frac{5}{2}{\gamma _2}\frac{1}{{{G_1}}} {\sin ^2}{i_{\rm{tot}}}\left( {\frac{{{G_1}}}{{{G_{\rm{tot}}}}}\frac{1}{{\sin {i_1}}} - \cot {i_1}} \right)\\
&\times e_1^2 \sin \left( {2{g_1}} \right)\frac{{{X_0}\sin {x_0}}}{{\sqrt {1 - X_0^2} }} + \frac{1}{4}{\gamma _2}\frac{1}{{{G_1}}}\sin \left( {2{i_{\rm tot}}} \right)\\
&\times \left( {2 + 3e_1^2 - 5e_1^2\cos 2{g_1}} \right)\left( {\cot {i_1} + \frac{{{X_0}\cos {x_0}}}{{\sqrt {1 - X_0^2} }}} \right),\\
{{\dot X}_0} =& \frac{5}{2}{\gamma _2}\frac{1}{{{G_1}}}e_1^2{\sin ^2}{i_{\rm{tot}}}\left( {\frac{{{G_1}}}{{{G_{\rm{tot}}}}}\frac{1}{{\sin {i_1}}} - \cot {i_1}} \right) \sin \left( {2{g_1}} \right)\\
&\times \sqrt {1 - X_0^2} \cos {x_0} - \frac{1}{4}{\gamma _2}\frac{1}{{{G_1}}}\sin \left( {2{i_{\rm tot}}} \right)\\
&\times \left( {2 + 3e_1^2 - 5e_1^2\cos 2{g_1}} \right)\sqrt {1 - X_0^2} \sin {x_0},
\end{aligned}
\end{equation*}
where the right-hand terms are time dependent because the orbital variables of the inner binary are periodic functions of time.

\begin{figure*}
\centering
\includegraphics[width=2\columnwidth]{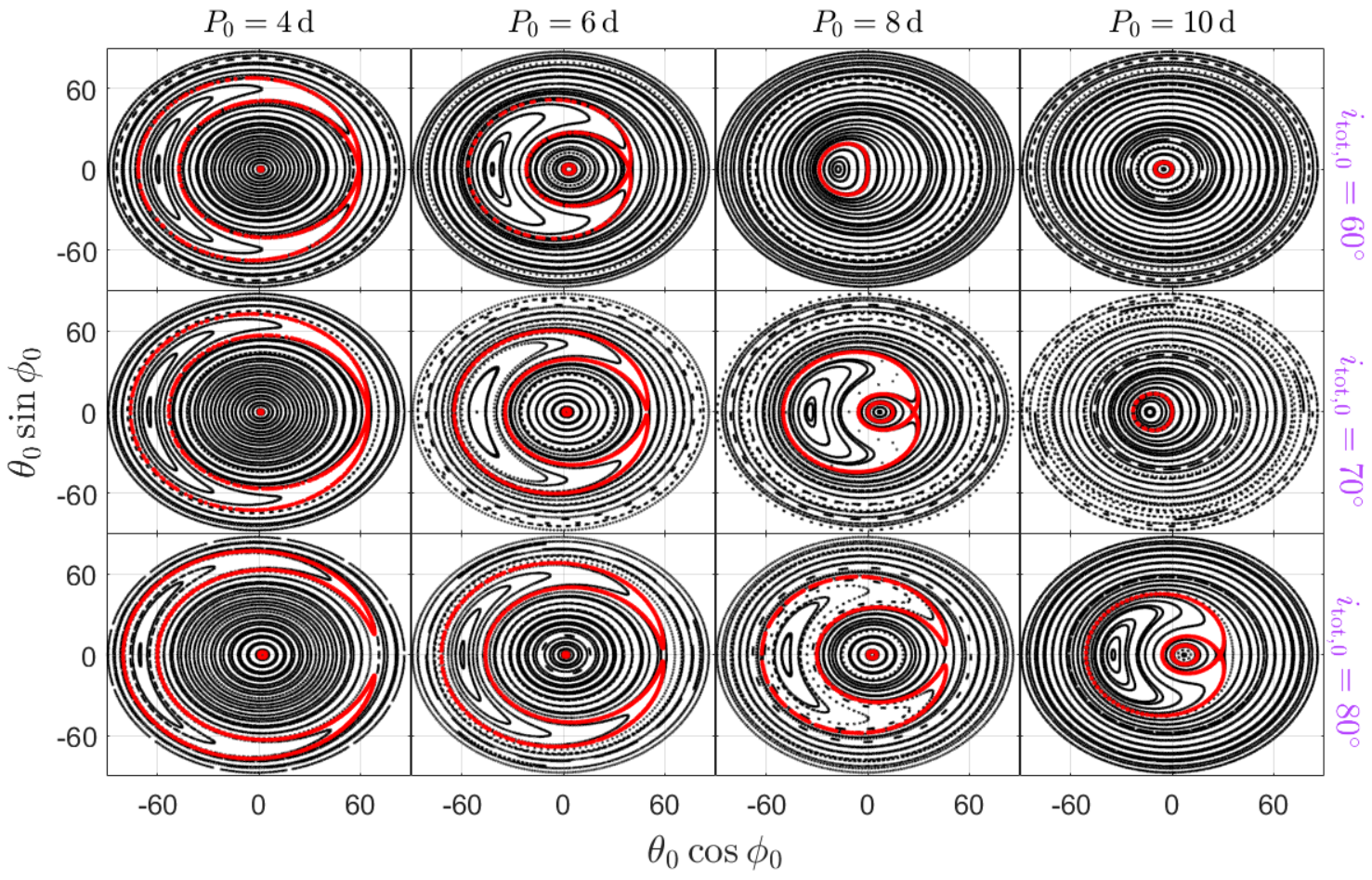}
\caption{Poincar\'e surfaces of section shown in the space of $(\theta_0\cos{\phi_0},\theta_0\sin{\phi_0})$ for the secular spin-orbit evolution of the primary star in circumbinary planetary systems with different spin periods $P_0$ and mutual inclination $i_{\rm tot,0}$. The argument of pericenter of the inner orbit is initially assumed at $\omega_{1,0}=0$. Please see text for the definition of Poincar\'e section. It should be mentioned that the sections corresponding to dynamical separatrices of librating islands are marked in red dots.}
\label{Fig5}
\end{figure*}

\begin{figure*}
\centering
\includegraphics[width=\columnwidth]{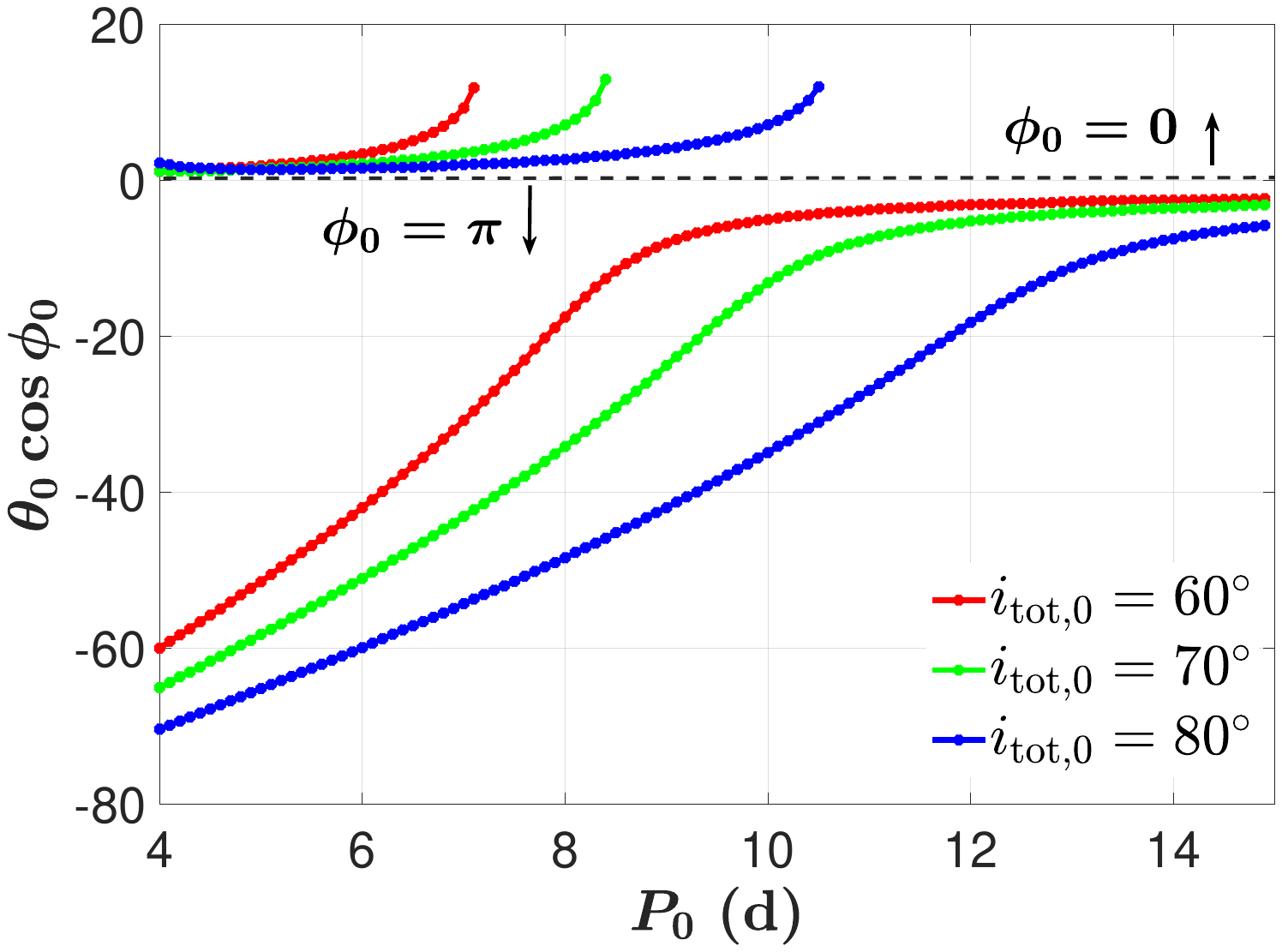}
\includegraphics[width=\columnwidth]{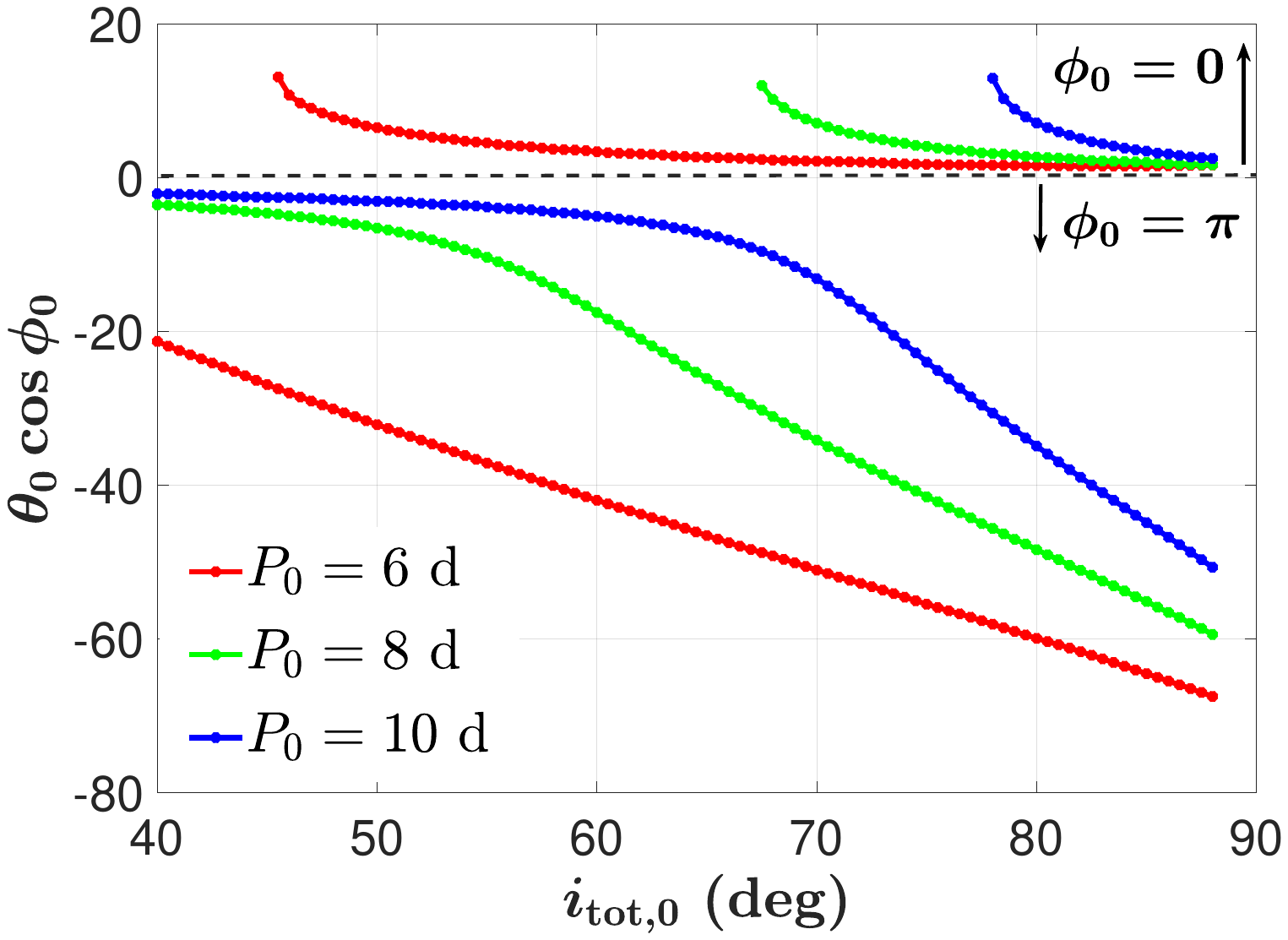}
\caption{Stable Cassini states (corresponding to stable equilibrium points at the centers of libration islands), numerically determined by analyzing Poincar\'e sections, for the stellar spin evolution in circumbinary planetary systems as a function of the spin period $P_0$ (\textit{the left panel}) and as a function of initial mutual inclination $i_{\rm tot,0}$ (\textit{the right panel}). It should be noted that the phase angle is located at $\phi_0 = 0$ for the families of Cassini state with $\theta_0\cos{\phi_0}>0$ and at $\phi_0 = \pi$ for the families of Cassini state with $\theta_0\cos{\phi_0}<0$. The unstable Cassini states are not considered. In the left panel, three levels of mutual inclinations at $i_{\rm tot,0}=60^{\circ}$, $70^{\circ}$ and $80^{\circ}$ are considered and, in the right panel, three values of spin periods at $P_0=6\,{\rm d}$, $8\,{\rm d}$ and $10\,{\rm d}$ are taken into account.}
\label{Fig6}
\end{figure*}

\section{Numerical exploration}
\label{Sect4}

Considering the fact that the spin Hamiltonian model is non-integrable, we adopt a numerical approach (Poincar\'e surfaces of section) to explore phase-space structures and determine the Cassini states. To be consistent with the cases of Figures \ref{Fig2} and \ref{Fig3}, the initial argument of pericenter is assumed at $\omega_{1,0}=0$, meaning that the inner orbit is of circulation, as shown in Figure \ref{Fig4}. It should be noted that the case of libration can be studied in a similar manner.    

\subsection{Poincar\'e sections}
\label{Sect4-1}

The spin Hamiltonian given by equation (\ref{Eq13}) determines a time-dependent model, where the orbital variables including $e_1$, $i_1$, $i_{\rm tot}$ and $g_1$ are periodic functions of time, we could define the Poincar\'e section by\footnote{For the case of libration, it is better to define the section by $\mod{\left(2g_1,2\pi\right)} = \pi$, as indicated by Figure \ref{Fig4}.}
\begin{equation}\label{Eq14}
\mod{\left(2g_1,2\pi\right)} = 0.
\end{equation}
It means that the spin states $(\theta_0,\phi_0)$ are recorded every time when the pericenter argument of the inner orbit periodically returns to its initial state at $2g_{1,0}=0$.

Figure \ref{Fig5} shows Poincar\'e sections in the space of $(\theta_0\cos{\phi_0},\theta_0\sin{\phi_0})$ for different combinations of the rotational period of the star at $P_0=4,6,8,10\, {\rm d}$ and initial mutual inclination at $i_{{\rm tot},0}=60^{\circ},70^{\circ},80^{\circ}$. The center of librating island corresponds to the (stable) Cassini state, and the boundaries of librating islands are shown in red dots.

The phase-space structure is determined by $i_{{\rm tot},0}$ and $P_0$. In general, two types of motion are observed, including librating and circulating cycles. Particularly, we can see that (a) there are two branches of (stable) Cassini states, one is at $\phi_0 =0$ and the other one is at $\phi_0=\pi$; (b) the branch of $\phi_0=\pi$ always exists, while the one of $\phi_0=0$ may disappear as the spin period $P_0$ is increased; (c) at a given $i_{{\rm tot},0}$, the island around $\phi_0=\pi$ shrinks and the Cassini state with $\phi_0=\pi$ approaches to the origin as the rotational period $P_0$ increases; and (d) at a given $P_0$, the island around $\phi_0=\pi$ enlarges and the Cassini state with $\phi_0=\pi$ moves away from the origin when the initial mutual inclination $i_{{\rm tot},0}$ becomes larger. 

\subsection{Numerical Cassini states}
\label{Sect4-2}

When tidal dissipation is considered, the stellar spin axis will usually evolve towards the stable Cassini states under an assumption of adiabatic approximation \citep[e.g.,][]{correia2015stellar}. Thus, the distribution of Cassini states in the conservative Hamiltonian model can help us to predict the outcome of the tidal migration.

By analyzing the Poincar\'e sections, we can numerically determine the location of stable Cassini state. Figure \ref{Fig6} shows the families of Cassini states at $\phi_0=0$ and $\phi_0=\pi$ as a function of $P_0$ in the left panel and as a function of $i_{{\rm tot},0}$ in the right panel. It should be noted that the unstable Cassini state is not considered here. 

At a given $i_{{\rm tot},0}$, there is a critical value of $P_0$, above which the family of Cassini state at $\phi_0=0$ disappear (see the left panel of Figure \ref{Fig6}). The critical value of $P_0$ increases with the initial mutual inclination. Similarly, at a given $P_0$, there is also a critical value of $i_{{\rm tot},0}$, below which the family of Cassini sate at $\phi_0=0$ disappears (see the right panel of Figure \ref{Fig6}). The critical value of $i_{{\rm tot},0}$ increases with the spin period $P_0$. The family of Cassini states at $\phi_0=\pi$ can exist in the entire ranges of $P_0$ and $i_{{\rm tot},0}$, and in this family a larger $i_{{\rm tot},0}$ or a smaller $P_0$ corresponds to a higher obliquity of Cassini state (i.e., a higher $\theta_0$).

\begin{figure*}
\centering
\includegraphics[width=2\columnwidth]{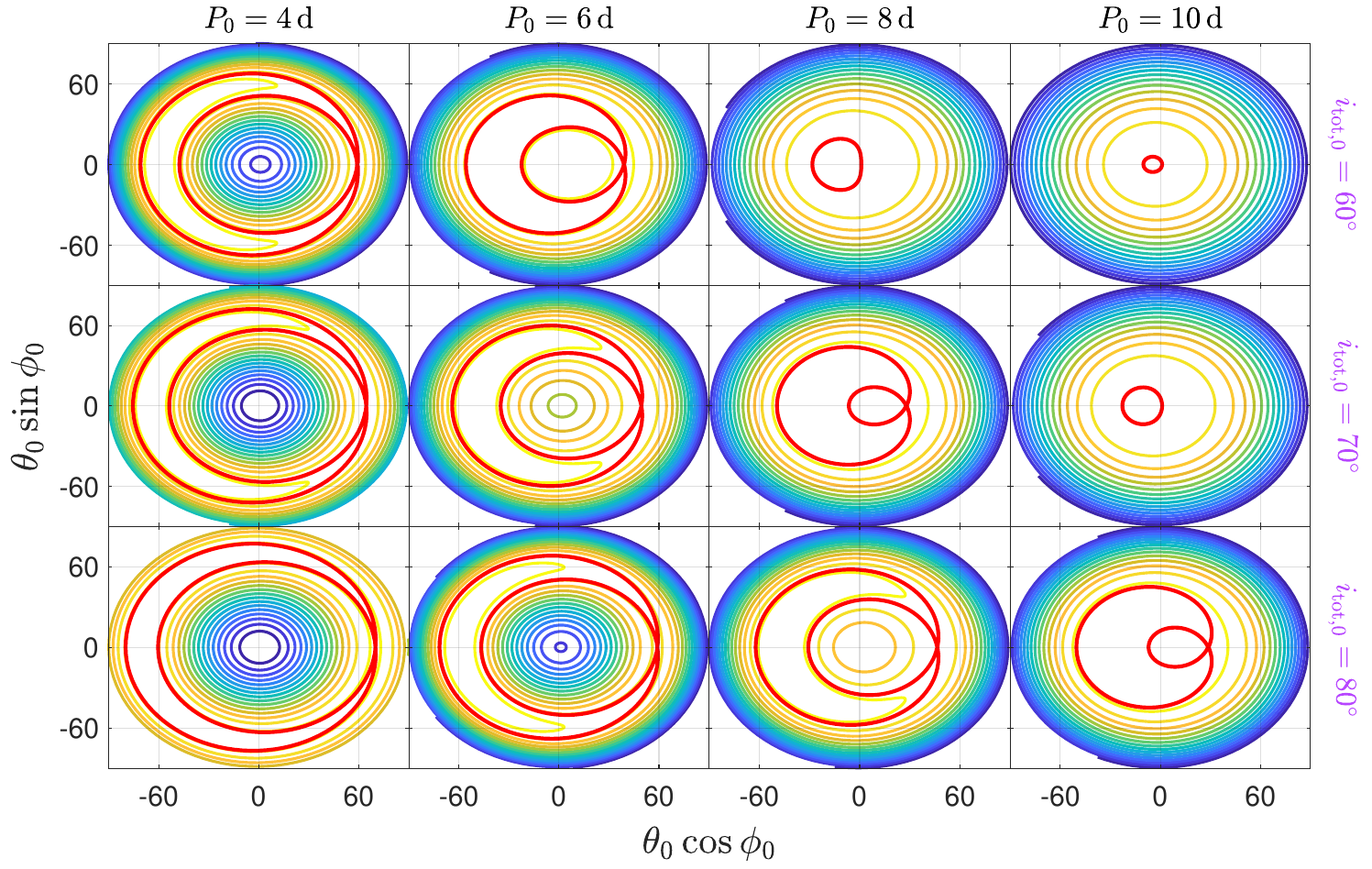}
\caption{Phase portraits shown in the space of $(\theta_0\cos{\phi_0},\theta_0\sin{\phi_0})$ for the stellar spin-orbit evolution in circumbinary planetary systems with spin periods at $P_0=4\,{\rm d}$, $6\,{\rm d}$, $8\,{\rm d}$ and $10\,{\rm d}$ and mutual inclinations at $i_{\rm tot,0}=60^{\circ}$, $70^{\circ}$ and $80^{\circ}$. Dynamical separatrices bounding librating islands centered at $\phi_0=\pi$ are shown in red lines.}
\label{Fig7}
\end{figure*}

\begin{figure*}
\centering
\includegraphics[width=\columnwidth]{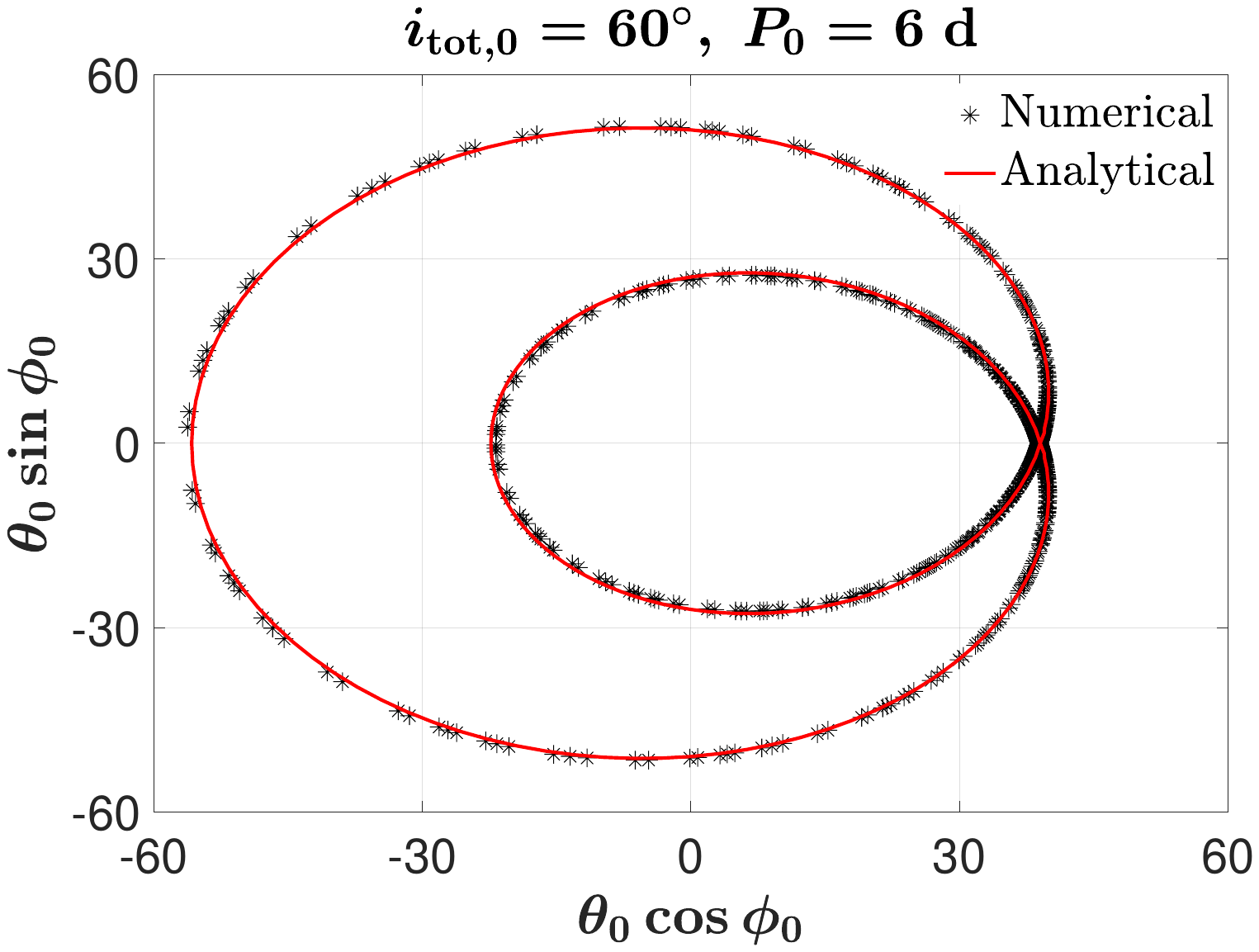}
\includegraphics[width=\columnwidth]{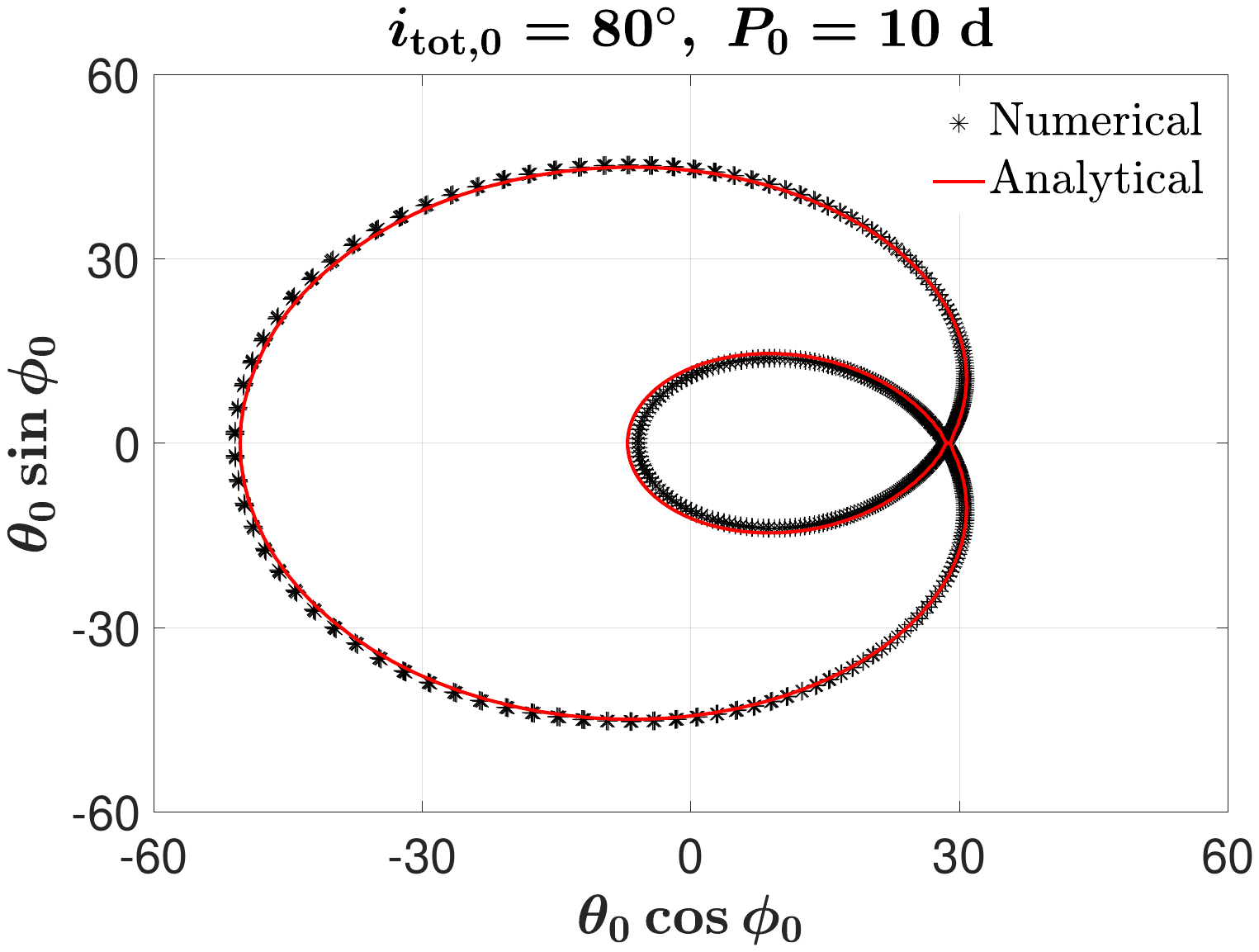}
\caption{Comparison of analytical and numerical separatrices in the cases of $i_{\rm tot,0}=60^{\circ}$, $P_0=6\,{\rm d}$ (\textit{left panel}), and $i_{\rm tot,0}=80^{\circ}$, $P_0=10\,{\rm d}$ (\textit{right panel}). The analytical separatrices come from the phase portraits and the numerical separatrices come from the associated Poincar\'e sections.}
\label{Fig8}
\end{figure*}

\section{Perturbative treatments}
\label{Sect5}

In this section, the phase-space structures explored in the previous section can be analytically reproduced from the viewpoint of perturbative treatment. 

According to Section \ref{Sect3}, it is known that the spin Hamiltonian given by equation (\ref{Eq13}) is closely dependent on the orbital evolution, which is governed by the quadrupole-order Hamiltonian given by equation (\ref{Eq4}). In particular, the orbit Hamiltonian determines an integrable model, where the orbital variables $e_1$, $i_1$ and $g_1$ are periodic functions of time. Denote the period of $g_1$ as $T$. This means that the coefficients of the spin Hamiltonian given by equation (\ref{Eq13}) are also periodic functions of time with period of $T$.

Because we concentrate on the secular spin evolution outside the spin-orbit resonances, we perform an average of the spin Hamiltonian over an orbital period of $g_1$ by
\begin{equation}\label{Eq15}
\left\langle {{{\cal H}_{\rm{spin}}}} \right\rangle  = \frac{1}{T}\int\limits_0^T {{{\cal H}_{\rm{spin}}}\left( {{g_1},{x_0},{X_0}} \right){\rm d}t},
\end{equation}
which can be equivalently written as
\begin{equation}\label{Eq16}
\left\langle {{{\cal H}_{\rm{spin}}}} \right\rangle  = \frac{1}{{2\pi }}\int\limits_0^{2\pi } {{{\cal H}_{\rm{spin}}}\left( {{g_1},{x_0},{X_0}} \right)\frac{1}{{{{\dot g}_1}}}\rm{d}{\rm{g}_1}},
\end{equation}
where the expression of ${\dot g}_1$ is provided by equation (\ref{Eq7}). It is difficult to derive an explicit expression for $\left\langle {{{\cal H}_{\rm{spin}}}} \right\rangle$. In practice, we compute the averaged spin Hamiltonian by numerical quadrature. 

Under the averaging approximation, the resulting averaged Hamiltonian $\left\langle {{{\cal H}_{\rm{spin}}}} \right\rangle$ is no longer time-dependent, and it determines an integrable dynamical model with $(x_0,X_0)$ as the pair of canonical variables. During the secular spin evolution, the Hamiltonian $\left\langle {{{\cal H}_{\rm{spin}}}} \right\rangle$ remains constant. As a result, secular evolution of stellar spin can be analyzed by plotting level curves of $\left\langle {{{\cal H}_{\rm{spin}}}} \right\rangle$ in the space of $(x_0,X_0)$. This corresponds to the so-called phase portraits.

Figure \ref{Fig7} shows phase portraits (level curves of the averaged spin Hamiltonian) in the space of $(\theta_0\cos{\phi_0},\theta_0\sin{\phi_0})$ with different combinations of spin period at $P_0=4, 6, 8, 10\,{\rm d}$ and initial mutual inclination at $i_{{\rm tot},0}=60^{\circ}, 70^{\circ}, 80^{\circ}$. In particular, dynamical separatrices of the librating islands around $\phi_0=\pi$ are marked by red lines. It is noted that the separatrices of the islands around $\phi_0=0$ are not shown here to avoid confusion. We can see that there is an excellent correspondence between phase portraits shown here and the Poincar\'e sections presented in Figure \ref{Fig5}. To validate the analytical treatment, we make a direct comparison about the analytical and numerical separatrices for two typical examples, as shown in Figure \ref{Fig8}. It further shows that the dynamical structures can be accurately captured by phase portraits under the averaged spin Hamiltonian.

\begin{figure*}
\centering
\includegraphics[width=\columnwidth]{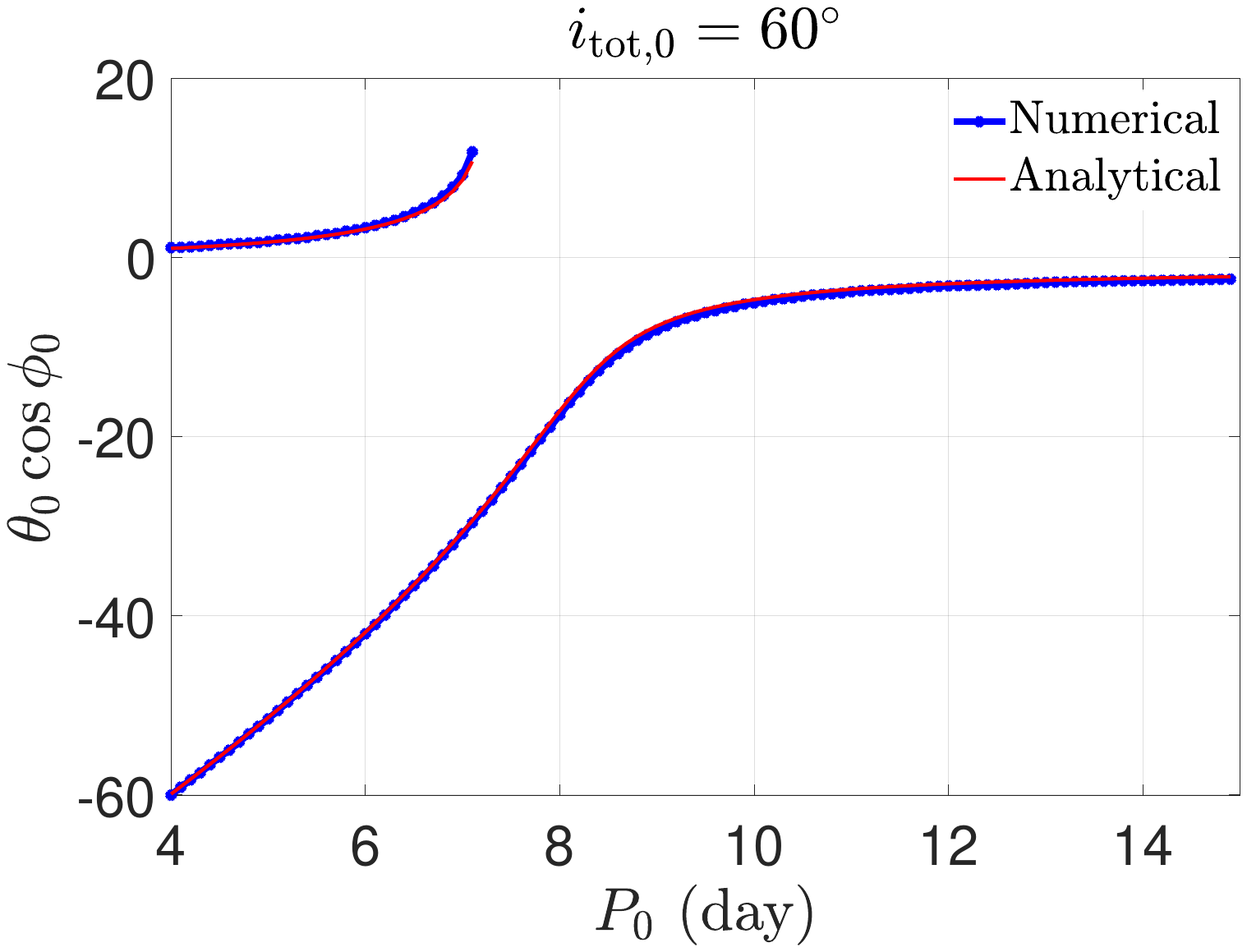}
\includegraphics[width=\columnwidth]{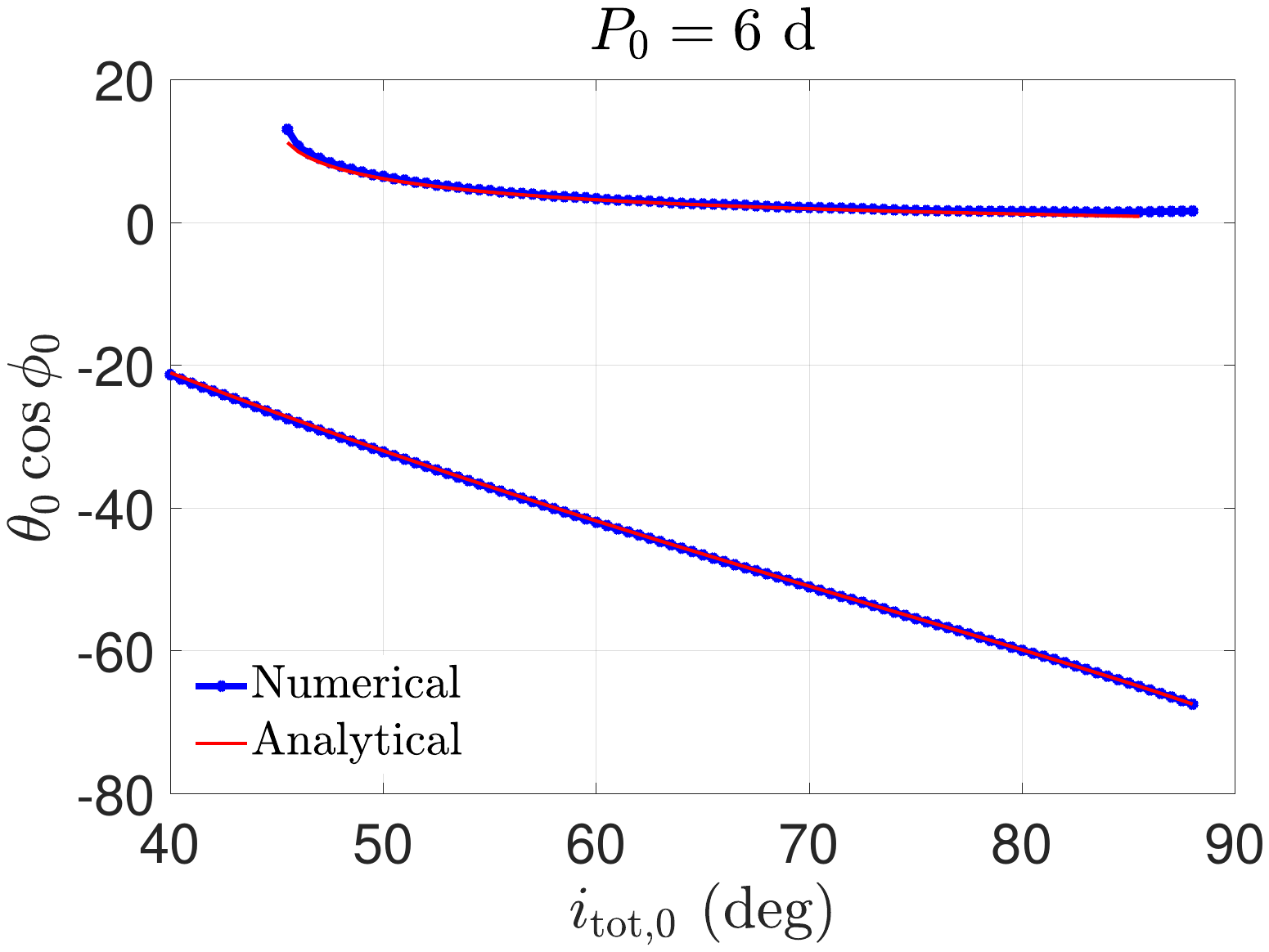}\\
\includegraphics[width=\columnwidth]{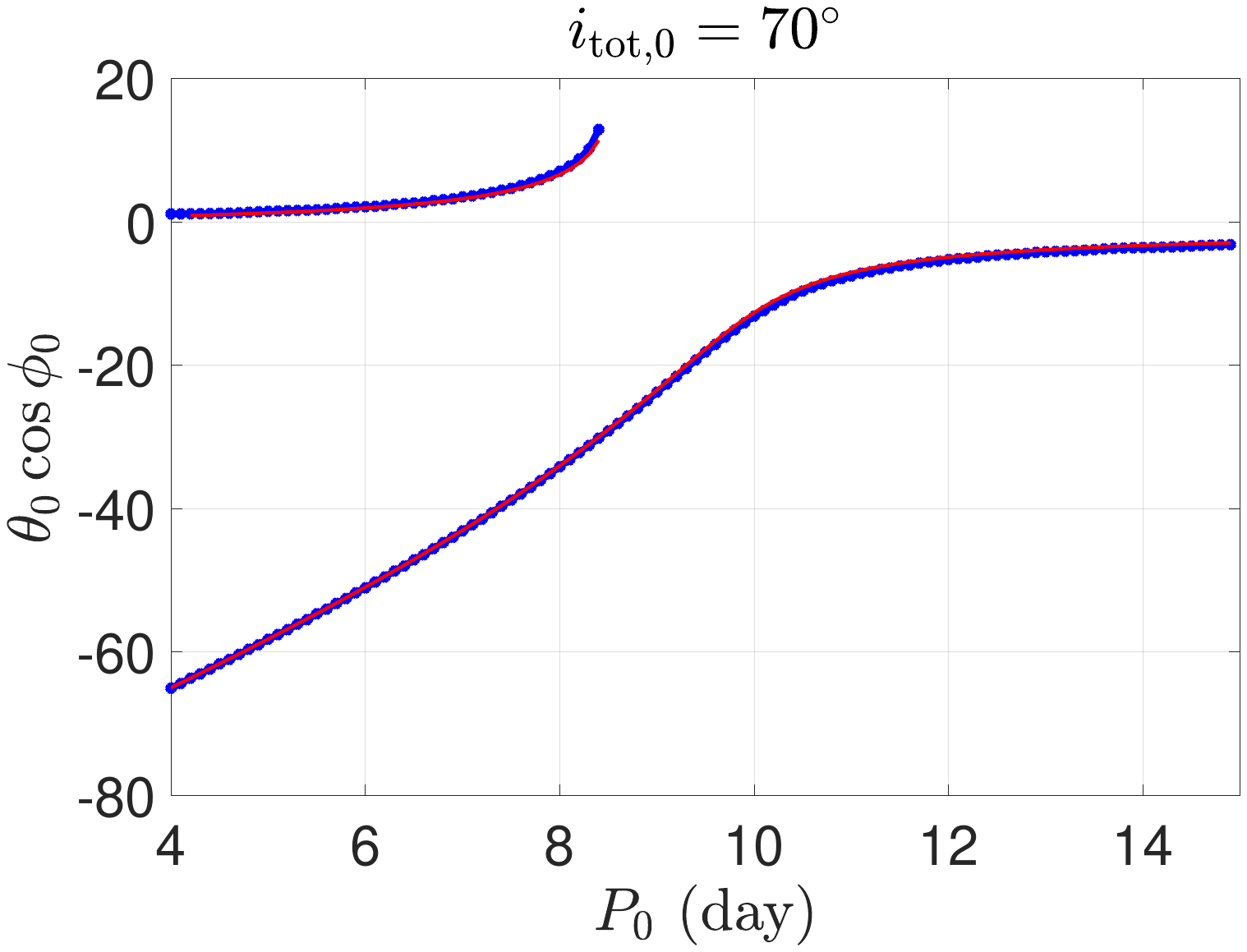}
\includegraphics[width=\columnwidth]{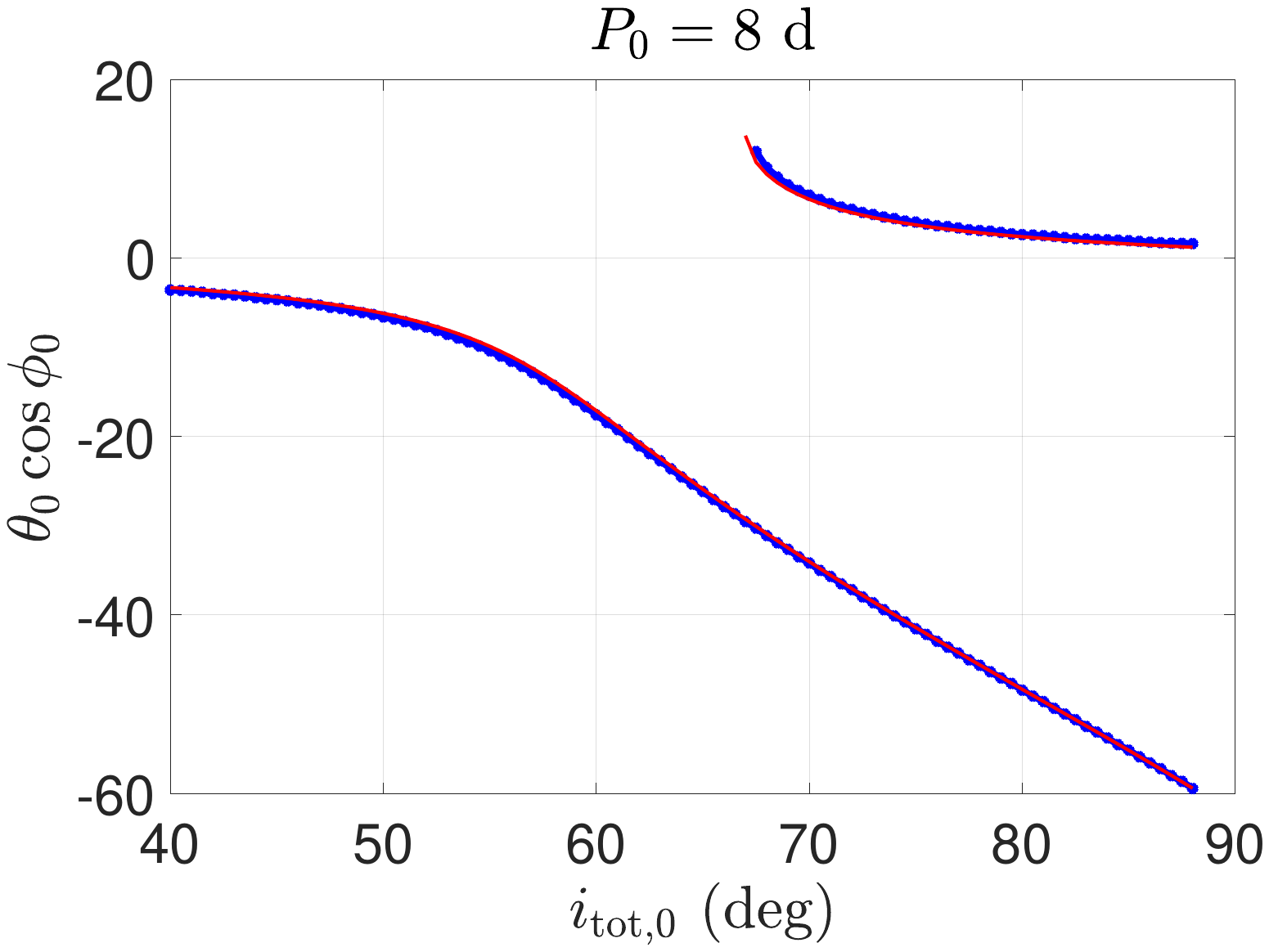}\\
\includegraphics[width=\columnwidth]{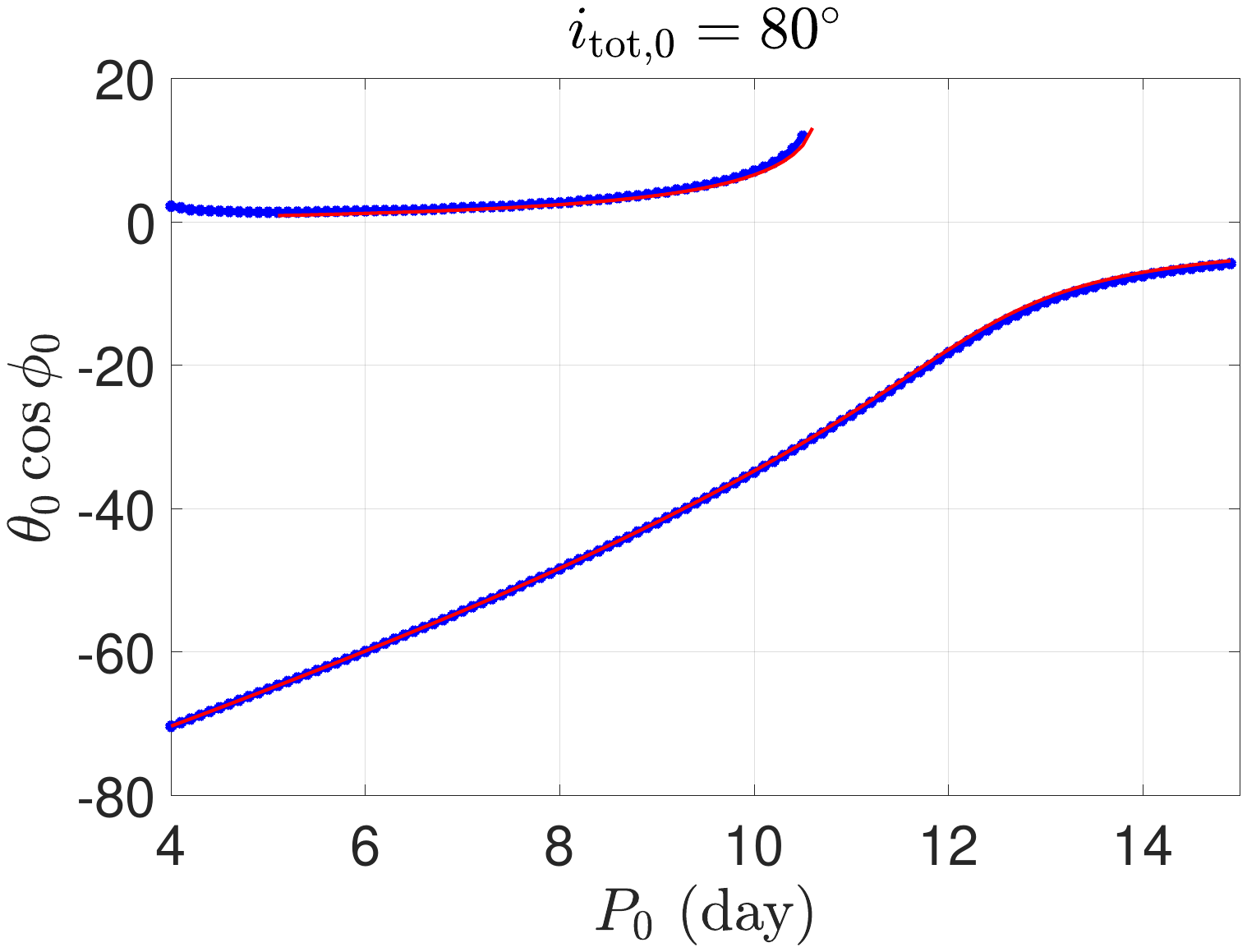}
\includegraphics[width=\columnwidth]{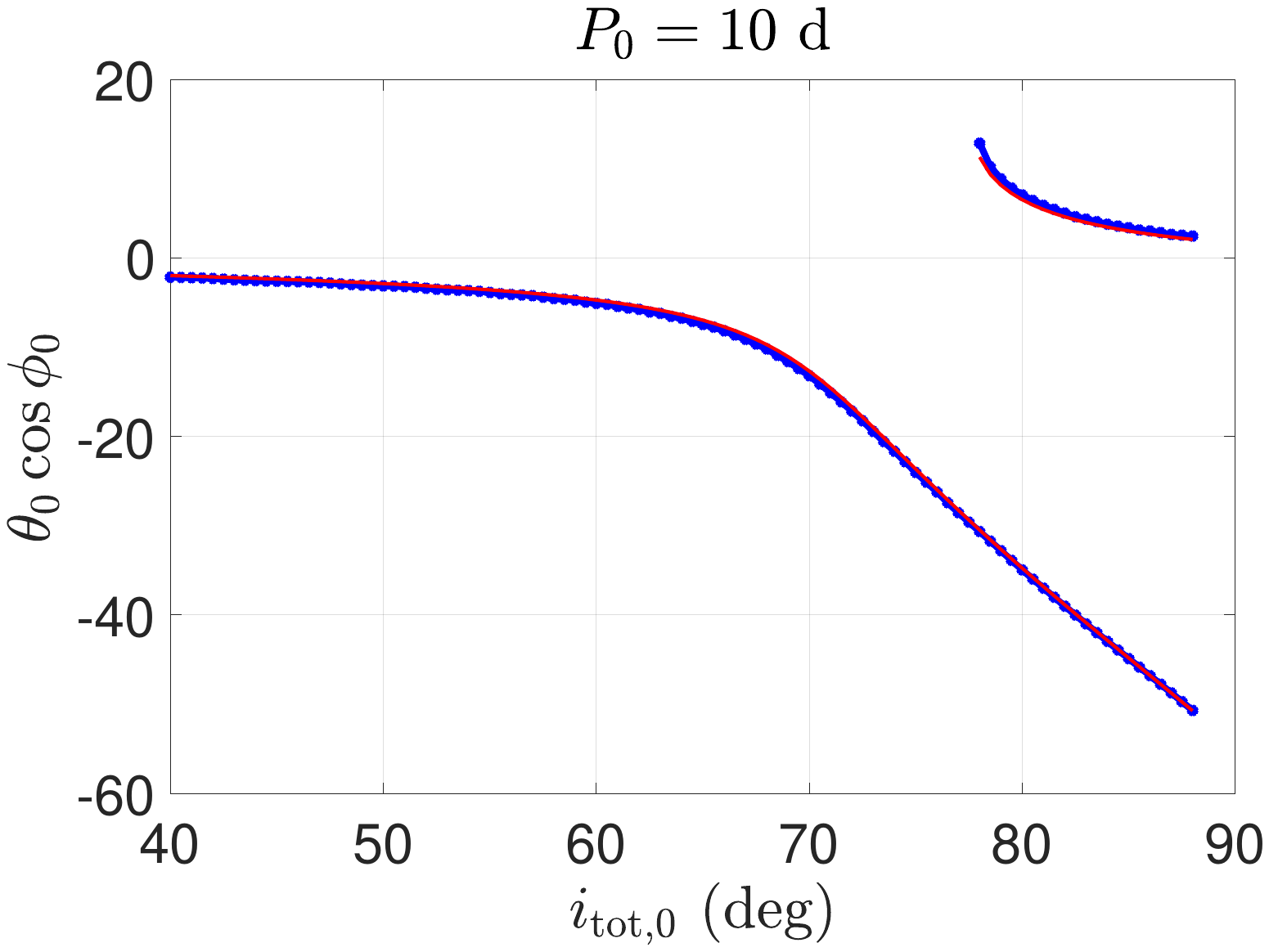}
\caption{Comparison of analytical and numerical (stable) Cassini states for initial mutual inclinations at $i_{\rm tot,0}=60^{\circ},70^{\circ},80^{\circ}$ (\textit{the left-column panels}) and for initial spin periods assumed at $P_0=6\,{\rm d}$, $8\,{\rm d}$ and $10\,{\rm d}$ (\textit{the right-column panels}). Similar to Figure \ref{Fig4}, the phase angle is located at $\phi_0 = 0$ for the families of Cassini state with $\theta_0\cos{\phi_0}>0$ and at $\phi_0 = \pi$ for the families of Cassini state with $\theta_0\cos{\phi_0}<0$. Particularly, the analytical Cassini states are determined by analyzing phase portraits, while the numerical Cassini states are produced from the associated Poincar\'e sections.}
\label{Fig9}
\end{figure*}

By analyzing phase portraits, we could determine the distribution of analytical Cassini states (center of librating islands). The analytical and numerical Cassini states are shown in Figure \ref{Fig9}, which shows that there is an excellent agreement between them. It is noted that the curves of numerical Cassini states are the same as the ones shown in Figure \ref{Fig6}.

One remark is made here. Under the orbital Hamiltonian model, the argument of pericenter ${\dot g}_1$ is, in general, not a constant, showing that the angle $g_1$ is a nonlinear function of time. Thus, the average of spin Hamiltonian over time is usually not equal to the average over $g_1$. Notice that the average of Hamiltonian over $g_1$ was taken in \citet{correia2015stellar} and \citet{correia2016secular}. According to Figure \ref{Fig4}, we can see that the nonlinearity of $g_1$ increases with the mutual inclination $i_{\rm tot}$. Thus, we can say that, only when the mutual inclination is low (i.e., nearly coplanar configuration), ${\dot g}_1$ can be assumed as a constant and only in this limit the average of spin Hamiltonian over $g_1$ is a good approximation.

\section{Applications to dissipative systems}
\label{Sect6}

The dynamics of stellar spin under the conservative system by turning off tidal effects are numerically studied in Section \ref{Sect4} by taking advantage of Poincar\'e sections and analytically studied in Section \ref{Sect5} by means of perturbative treatments. In particular, phase-space structures are produced, and the families of stable Cassini states are determined. There is an excellent agreement between the analytical and numerical results. In this section, the analytical results are applied to understanding the coupled evolution of mutual inclination $i_{\rm tot}$ and stellar obliquity $\theta_0$ in the presence of tidal dissipation.

\begin{figure*}
\centering
\includegraphics[width=\columnwidth]{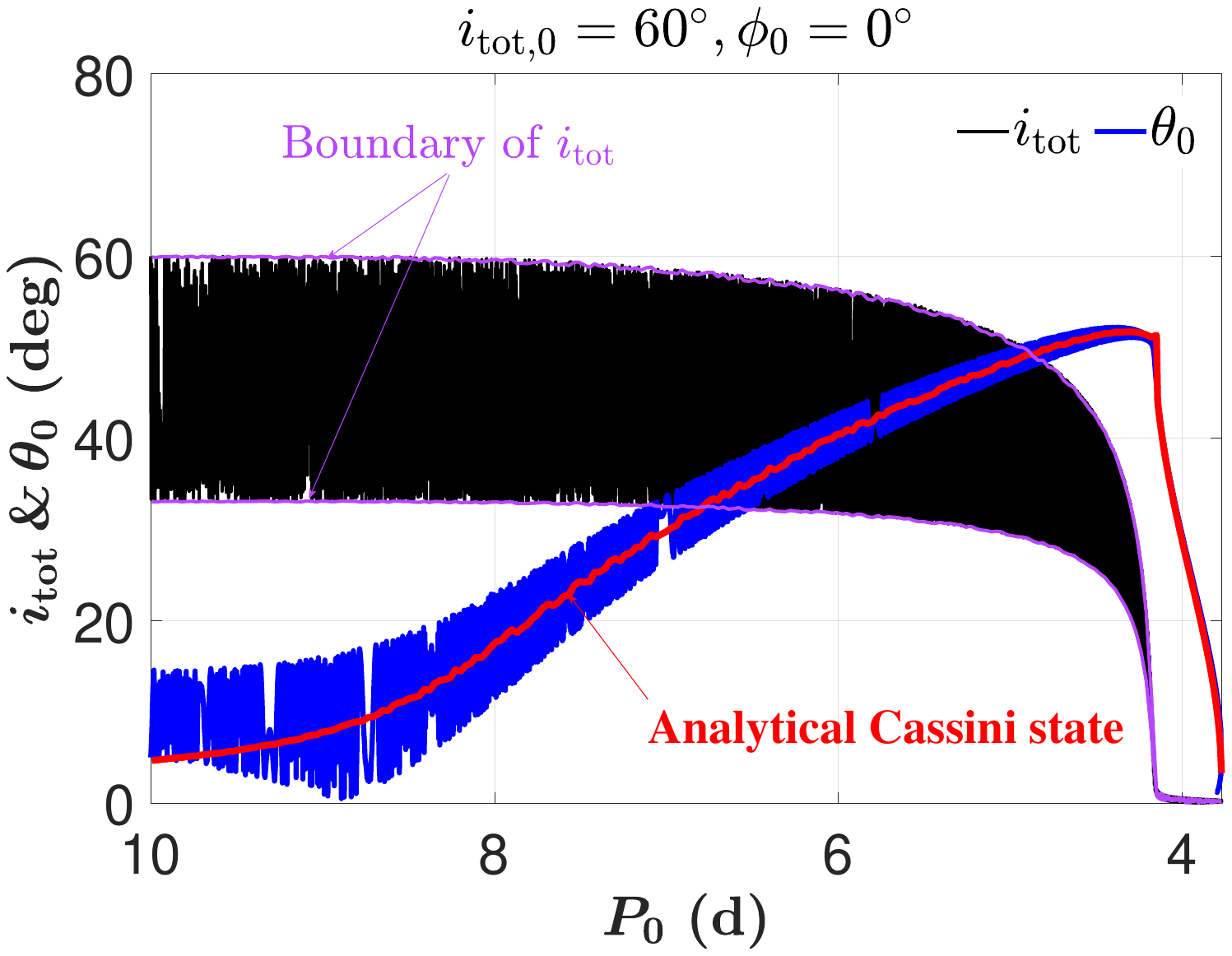}
\includegraphics[width=\columnwidth]{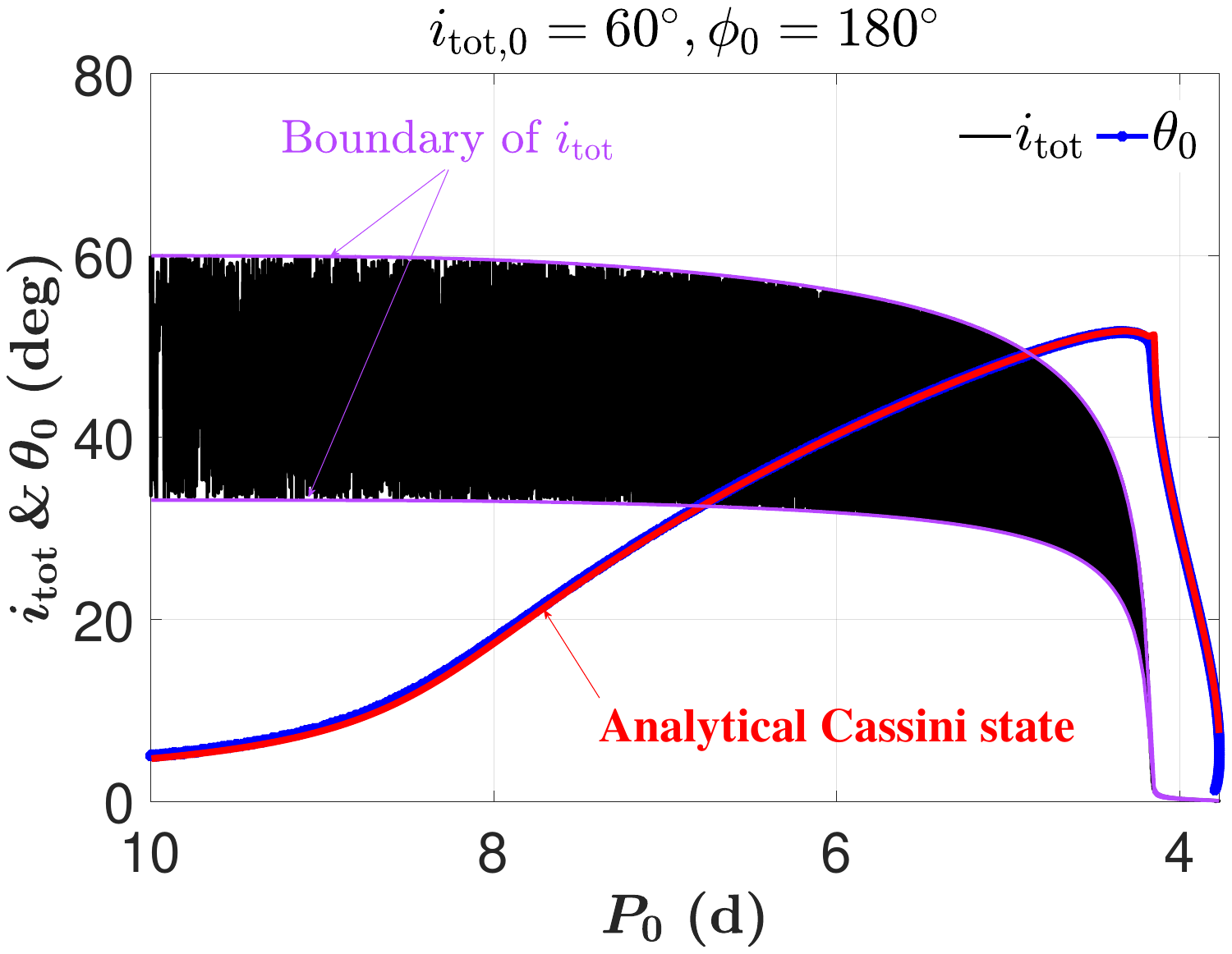}\\
\includegraphics[width=\columnwidth]{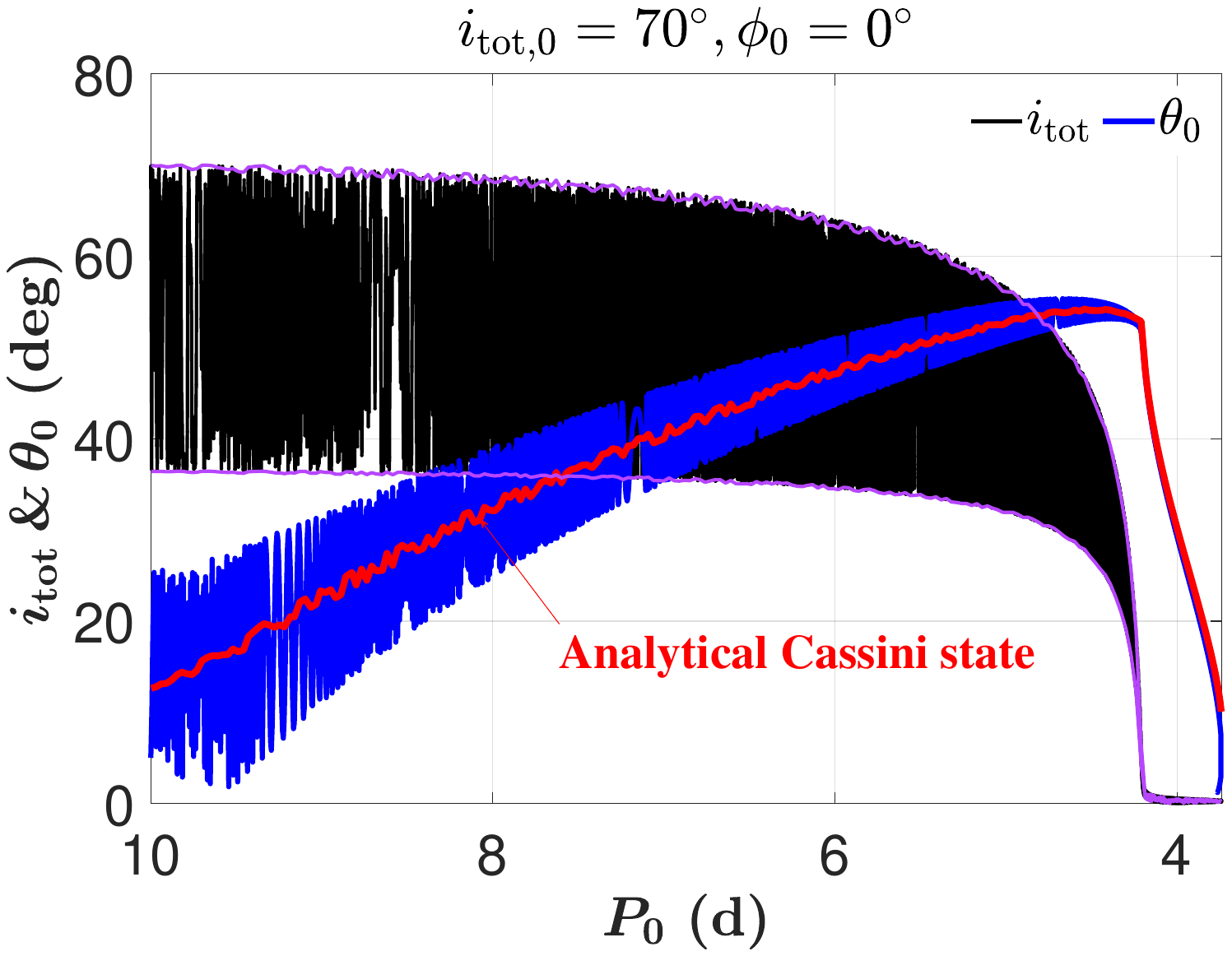}
\includegraphics[width=\columnwidth]{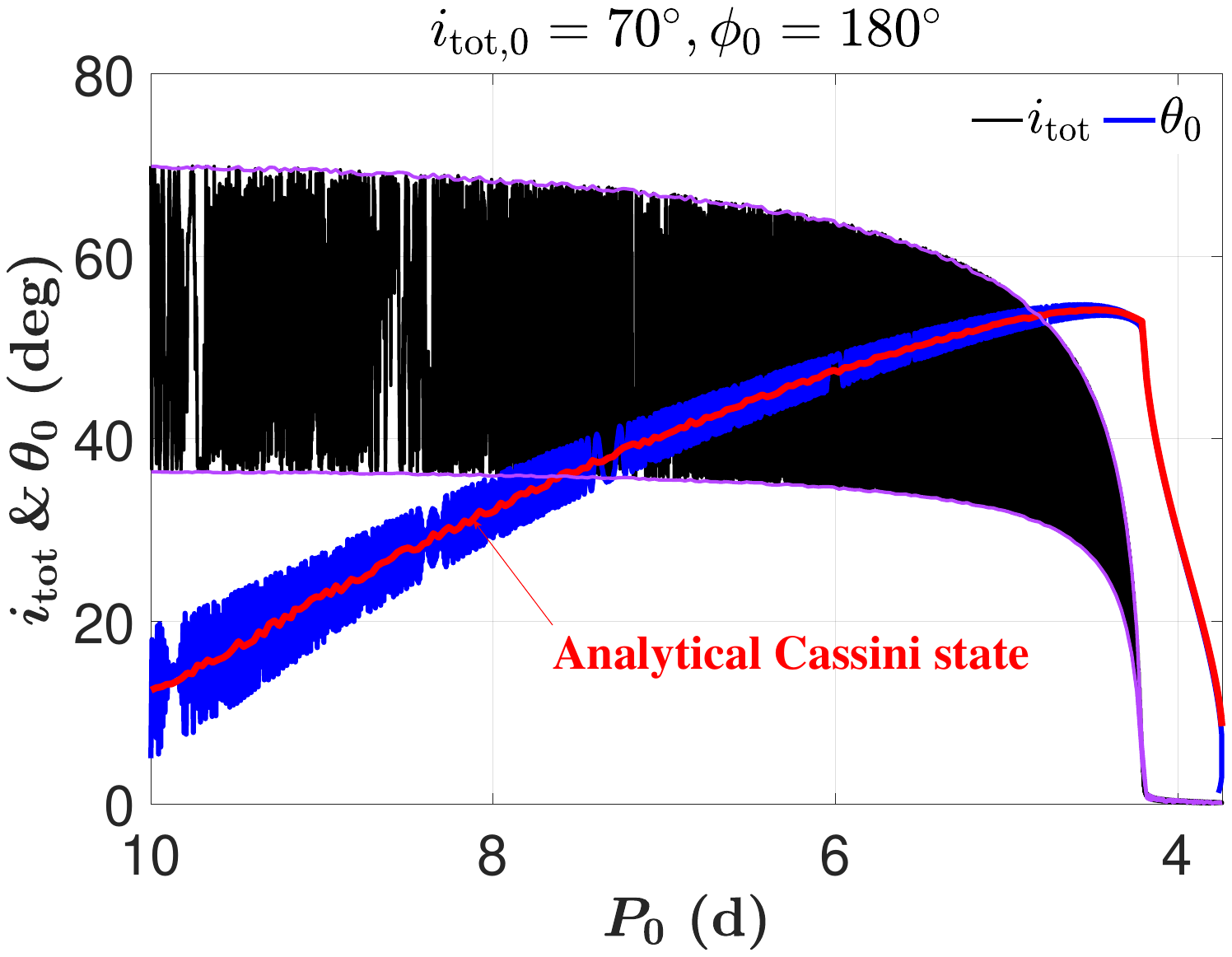}\\
\includegraphics[width=\columnwidth]{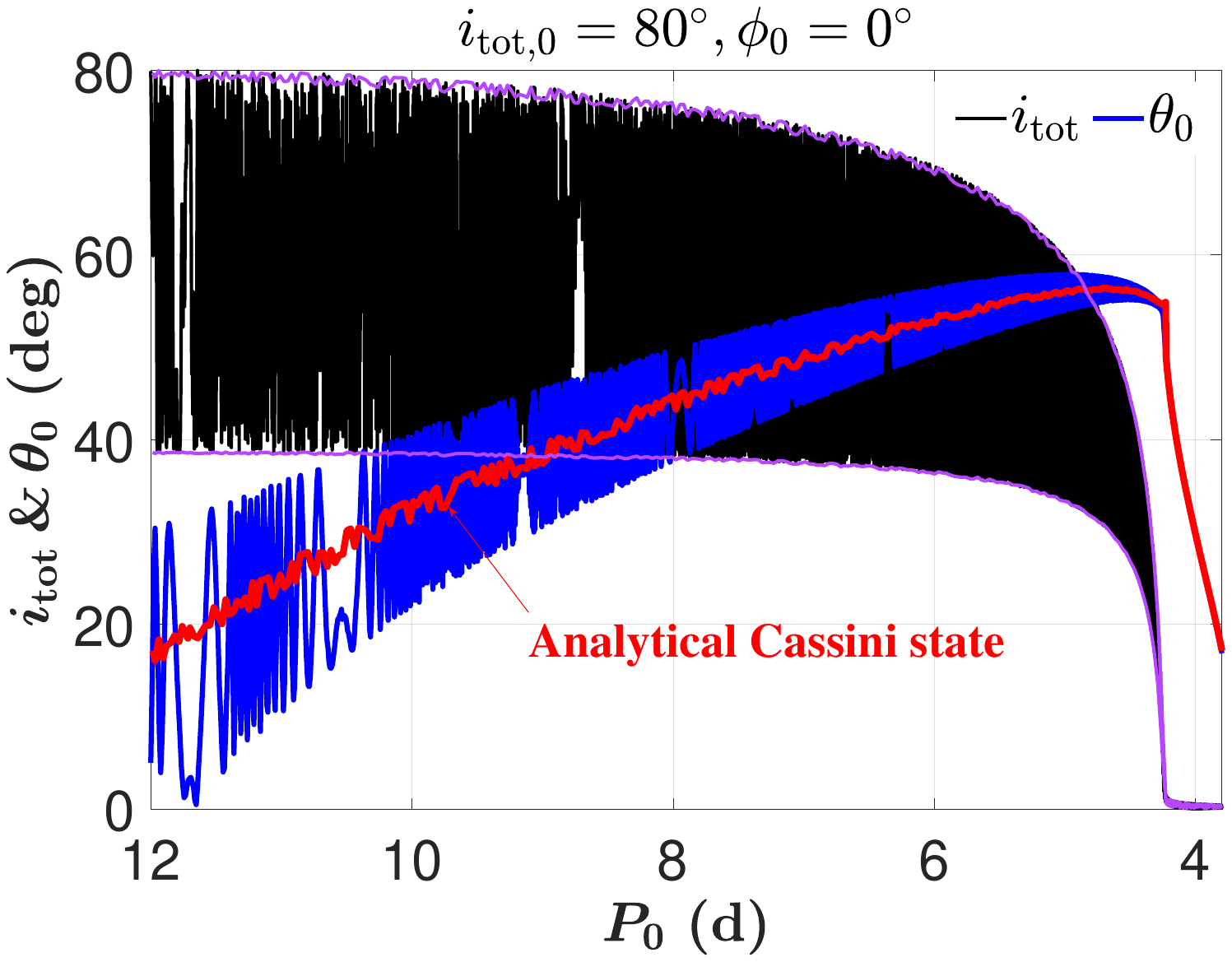}
\includegraphics[width=\columnwidth]{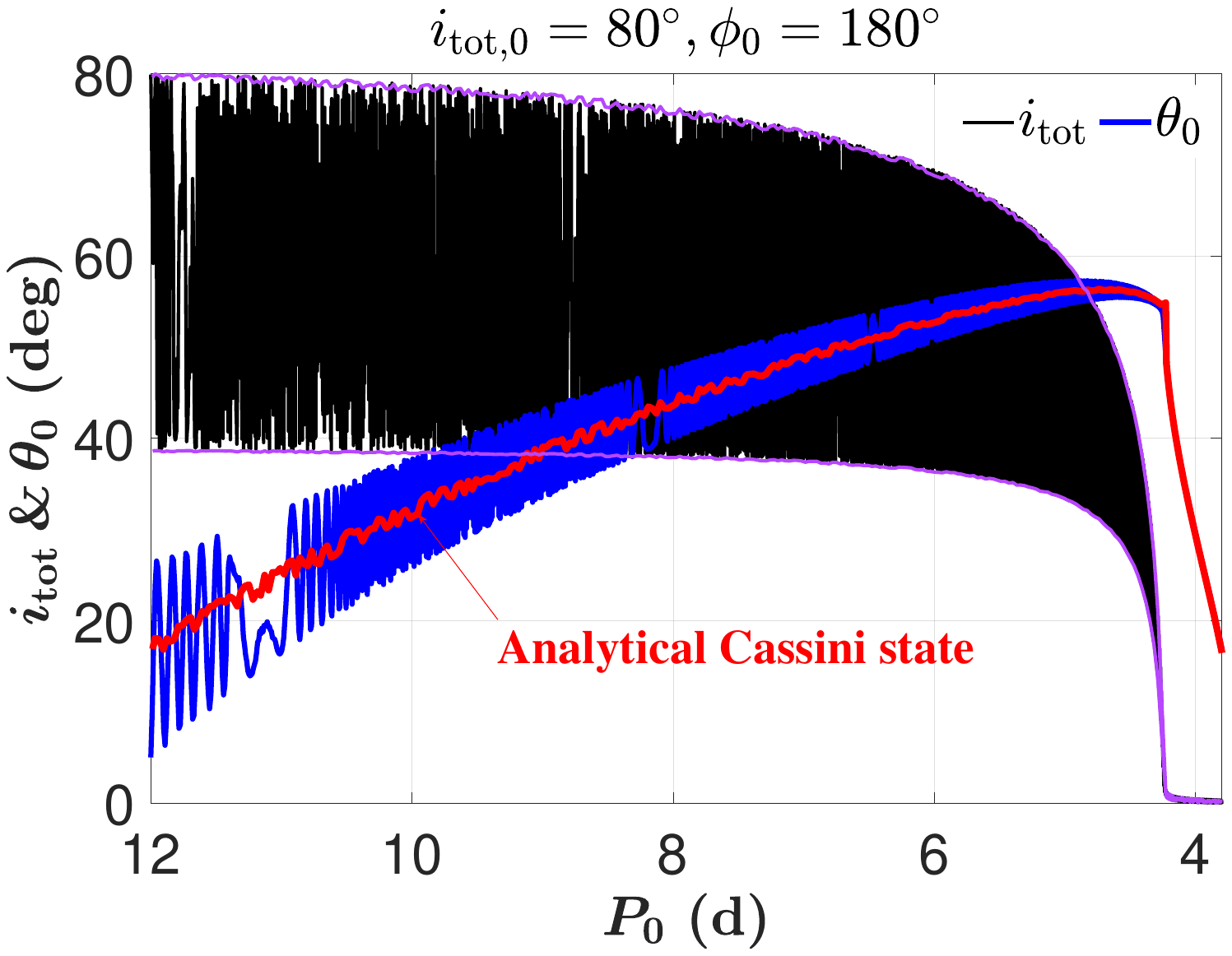}
\caption{Analytical Cassini states evaluated at $\phi_0=\pi$ as functions of the stellar rotation period $P_0$ (red lines), together with numerical evolutions of mutual inclination $i_{\rm tot}$ and stellar obliquity $\theta_0$, for circumbinary planetary systems with initial mutual inclinations at $i_{\rm tot,0}=60^{\circ}$ (\textit{top-row panels}), $70^{\circ}$ (\textit{middle-row panels}) and $80^{\circ}$ (\textit{bottom-row panels}). The stellar obliquity is initially assumed at $\theta_0=5^{\circ}$, and the initial phase angle of stellar spin is assumed at $\phi_0 = 0$ for the left-column panels and it is at $\phi_0=\pi$ for the right-column ones. The boundaries of the mutual inclination are evaluated from the quadrupole-order Hamiltonian during the tidal evolution.}
\label{Fig10}
\end{figure*}

The simplified model with tidal effects, discussed in Section \ref{Sect2}, is adopted. In this model, both the secondary star $m_1$ and the faraway planet $m_2$ are considered as point-mass bodies (i.e., assuming $k_{21}=k_{22}=0$) and the third-body perturbation is truncated at the quadrupole order. The secular spin-orbit evolution equations are numerically integrated over 3 Gyr. The parameters of circumbinary system are the same as those adopted in Figures \ref{Fig2} and \ref{Fig3}. During the tidal evolution, the spin period of the primary star $P_0$ decreases with time and it evolves towards a synchronous configuration. For convenience, the evolutions of $i_{\rm tot}$ and $\theta_0$ are shown in Figure \ref{Fig10} as functions of the stellar spin period $P_0$, where the cases of $i_{{\rm tot}, 0} = 60^{\circ}, 70^{\circ}, 80^{\circ}$ with $\phi_0 = 0$ and $\phi_0 = \pi$ are taken into account. In particular, the curves of $i_{\rm tot}$ are shown in black lines and the curves of $\theta_0$ are shown by blue lines. Please refer to the caption for the setting of initial conditions. It is noted that, during the tidal evolution, both the semimajor axis and eccentricity of the inner binary $(a_1,e_1)$ decrease with time (see Figure \ref{Fig2} for example). Additionally, the semimajor axis and eccentricity of the outer binary $(a_2,e_2)$ exhibit negligible variations during the tidal evolution. To save space, the evolutions of $a_{1,2}$ and $e_{1,2}$ are not presented here. 

From Figure \ref{Fig10}, we can see that (a) in all cases there is a coupled evolution of mutual inclination $i_{\rm tot}$ and stellar obliquity $\theta_0$; (b) the mutual inclination $i_{\rm tot}$ is damped with a decreasing amplitude of oscillation; (c) in the cases of $\phi_0=\pi$ the stellar obliquity $\theta_0$ first decreases, then increases, and finally decreases towards zero on average; and (d) in the cases of $\phi_0=0$ the stellar obliquity $\theta_0$ first increases, and then decreases towards zero on average.

What is the dynamical mechanism underlying the coupled evolution of $i_{\rm tot}$ and $\theta_0$? 

It is known in Section \ref{Sect2} that the secular and tidal evolution is an approximately adiabatic process. As a result, at an arbitrary moment of time, we can approximate the dynamics from the viewpoint of conservative system. 

Firstly, the orbital evolution can be approximated by the integrable quadrupole-order Hamiltonian, as shown by the phase portrait in Figure \ref{Fig4}. As a result, the maximum and minimum values of $i_{\rm tot}$ within an orbital period of $g_1$ can be analytically determined. See the purple lines in Figure \ref{Fig10}. 

Secondly, at an arbitrary moment of time, the Cassini states can be analytically determined. Please see the red lines shown in Figure \ref{Fig10}. According to orbital configurations, there are two stages of evolution for stellar obliquity. In the first stage, the orbital configuration is inclined, and in this case the stellar obliquity $\theta_0$ evolves around the Cassini state with a decreasing amplitude of oscillation. In the second stage, the orbital configuration becomes coplanar, and in this case the short-period oscillation of stellar obliquity $\theta_0$ vanishes. In particular, the mutual inclination can be approximated as zero and the evolution of stellar obliquity $\theta_0$ is subject to the conservation of the total angular momentum, given by
\begin{equation}\label{Eq17}
\cos {\theta _0} = \frac{{{G^2} - L_0^2 - {{\left( {{G_1} + {G_2}} \right)}^2}}}{{2{L_0}\left( {{G_1} + {G_2}} \right)}},
\end{equation}
where $G$ is the total angular momentum, determined by the initial condition.

The coupled evolution of $i_{\rm tot}$ and $\theta_0$ in dissipative systems can be well understood with the aid of adiabatic approximation. In particular, the damping of the mutual inclination is a dynamical outcome of the angular momentum exchange between the spin of the primary star and the orbit of the planet under the conservation of the total angular momentum. We can summarize that (a) the boundaries of $i_{\rm tot}$ derived from the quadrupole-order orbit Hamiltonian can provide good envelope for the oscillation of $i_{\rm tot}$ during the tidal evolution; (b) in the stage of non-planar configuration the stellar obliquity $\theta_0$ evolves along the curve of Cassini states with decreasing amplitude of short-period oscillations; and (c) in the stage of co-planar configuration the stellar obliquity $\theta_0$ follows closely along the curve of total angular momentum of circumbinary system.

\section{Conclusions}
\label{Sect7}

In this work, we study the secular and tidal evolution of circumbinary planetary systems within the non-restricted dynamical framework developed by \citet{correia2016secular}. This model includes the octupole-order approximation for third-body perturbations, general relativistic corrections, rotational and tidal deformations, and tidal dissipation. Within this framework, a striking coupled evolutionary behavior is observed: an initially arbitrary planetary inclination can be damped to zero, while the stellar obliquity is first excited to a high level and subsequently decreases toward zero. We systematically investigate the dynamical mechanism responsible for this coupled evolution.

We first numerically integrate the complete dynamical model for a representative example considered in \citet{correia2016secular}, reproducing their Figure 15, which shows the coupled evolution of the mutual inclination $i_{\rm tot}$ and the stellar obliquity $\theta_0$. We then simplify the model by (a) truncating the third-body perturbation at the quadrupole order and (b) treating the secondary star and the planet as point masses. The simplified model successfully reproduces the coupled evolution of $i_{\rm tot}$ and $\theta_0$.

With tidal effects turned off, the orbital and spin Hamiltonian can be formulated independently. Under the quadrupole-order approximation, the orbital Hamiltonian defines an integrable model, leading to periodic evolution of the orbital elements of the inner and outer binaries. The spin Hamiltonian, however, is explicitly time dependent and corresponds to a 1.5 DOF system. To investigate the phase-space structure of the stellar spin, we construct Poincaré sections, which reveal two families of Cassini states located at $\phi_0=0$ and $\phi_0=\pi$. The distributions of stable Cassini states are then mapped in the parameter space of spin period $P_0$ and initial mutual inclination $i_{{\rm tot},0}$. The family at $\phi_0=\pi$ exists throughout the entire $(i_{{\rm tot},0}, P_0)$ space, and the associated stellar obliquity $\theta_0$ increases as the spin period $P_0$ decreases or the initial mutual inclination $i_{{\rm tot},0}$ increases. These numerical results, including phase-space structures and Cassini states, can be reproduced analytically using perturbative methods. Specifically, we average the spin Hamiltonian over one period of the inner orbit’s argument of pericenter $g_1$, which yields an integrable averaged model. Phase portraits obtained from the level curves of the averaged Hamiltonian show a clear correspondence with the Poincaré sections. Moreover, the Cassini states identified from the phase portraits agree remarkably well with those obtained numerically.

The results derived from the conservative Hamiltonian framework provide a natural interpretation for the coupled evolution of mutual inclination and stellar obliquity in the presence of tidal dissipation. During the non-planar stage, the mutual inclination steadily decreases due to tides, while the stellar obliquity evolves around Cassini states with a progressively shrinking oscillation amplitude due to adiabatic capture and conservation of total angular momentum. When the orbital configuration becomes coplanar, the stellar obliquity reaches its maximum value. The system subsequently enters the coplanar stage, in which tidal dissipation damps the stellar obliquity until spin-orbit alignment is attained. In particular, the evolution of stellar obliquity during this second stage closely follows the curve defined by the conservation of total angular momentum (see equation \ref{Eq17}).


\newpage
\appendix

\section{Secular evolution equations of spins and orbits}
\label{Appendix_A}

The orbital states of the inner and outer binaries are described by the orbital angular momentum vectors ${\bm G}_{1,2}$, and the spin states of the oblate bodies $m_{0,1,2}$ are described by the rotational angular momentum vectors ${\bm L}_{0,1,2}$. During the tidal evolution, the system energy is dissipated, while the total angular momentum is conserved. The tidal model can therefore be seen as a mathematical formulation that combines energy transformation with conservation of angular momentum. Following this rule, \citet{correia2016secular} developed an elegant formulism in a vectorial form for the averaged equations of motion under the non-restricted hierarchal three-body problem as follows: 
\begin{equation}\label{Eq_A1}
\begin{aligned}
{{\dot {\bm L}}_0} &= {\left( {{{\dot {\bm L}}_0}} \right)_{\rm tide}} + {\left( {{{\dot {\bm L}}_0}} \right)_{\rm rot}},\quad {{\dot {\bm L}}_1} = {\left( {{{\dot {\bm L}}_1}} \right)_{\rm tide}} + {\left( {{{\dot {\bm L}}_1}} \right)_{\rm rot}},\quad {{\dot {\bm L}}_2} = {\left( {{{\dot {\bm L}}_2}} \right)_{\rm tide}} + {\left( {{{\dot {\bm L}}_2}} \right)_{\rm rot}},\\
{{\dot {\bm G}}_1} &= {\left( {{{\dot {\bm G}}_1}} \right)_{\rm tide}} + {\left( {{{\dot {\bm G}}_1}} \right)_{\rm rot}} + {\left( {{{\dot {\bm G}}_1}} \right)_{\rm quad}} + {\left( {{{\dot {\bm G}}_1}} \right)_{\rm oct}},\quad {{\dot {\bm G}}_2} = {\left( {{{\dot {\bm G}}_2}} \right)_{\rm tide}} + {\left( {{{\dot {\bm G}}_2}} \right)_{\rm rot}} + {\left( {{{\dot {\bm G}}_2}} \right)_{\rm quad}} + {\left( {{{\dot {\bm G}}_2}} \right)_{\rm oct}},\\
{{\dot {\bm e}}_1} &= {\left( {{{\dot {\bm e}}_1}} \right)_{\rm tide}} + {\left( {{{\dot {\bm e}}_1}} \right)_{\rm rot}} + {\left( {{{\dot {\bm e}}_1}} \right)_{\rm quad}} + {\left( {{{\dot {\bm e}}_1}} \right)_{\rm oct}} + {\left( {{{\dot {\bm e}}_1}} \right)_{\rm gr}},\quad {{\dot {\bm e}}_2} = {\left( {{{\dot {\bm e}}_2}} \right)_{\rm tide}} + {\left( {{{\dot {\bm e}}_2}} \right)_{\rm rot}} + {\left( {{{\dot {\bm e}}_2}} \right)_{\rm quad}} + {\left( {{{\dot {\bm e}}_2}} \right)_{\rm oct}},
\end{aligned}
\end{equation}
which considers the octupole-order approximation for the third-body perturbation (with subscripts `quad' for the quadrupole-order approximation and `oct' for the octupole-order approximation), general relativity corrections (with subscript `gr'), quadrupole-order approximation for rotation deformations (with subscript `rot') and tidal effects (with subscript `tide'). The notation system adopted in this work is the same as that of \citet{correia2016secular}. For self-consistent purpose, here we provide the detailed expression of each component.

The quadrupole-order contribution of the third-body perturbation is given by
\begin{equation*}
\begin{aligned}
{\left( {{{\dot {\bm G}}_1}} \right)_{\rm quad}} =&  - {\gamma _2}\left[ {\left( {1 - e_1^2} \right)\left( {{{\hat {\bm k}}_1} \cdot {{\hat {\bm k}}_2}} \right){{\hat {\bm k}}_2} \times {{\hat {\bm k}}_1} - 5\left( {{{\bm e}_1} \cdot {{\hat {\bm k}}_2}} \right){{\hat {\bm k}}_2} \times {{\bm e}_1}} \right],\quad {\left( {{{\dot {\bm G}}_2}} \right)_{\rm quad}} =  - {\left( {{{\dot {\bm G}}_1}} \right)_{\rm quad}},\\
{\left( {{{\dot {\bm e}}_1}} \right)_{\rm quad}} =&  - \frac{{{\gamma _2}\left( {1 - e_1^2} \right)}}{{\left\| {{{\bm G}_1}} \right\|}}\left[ {\left( {{{\hat {\bm k}}_1} \cdot {{\hat {\bm k}}_2}} \right){{\hat {\bm k}}_2} \times {{\bm e}_1} - 2{{\hat {\bm k}}_1} \times {{\bm e}_1} - 5\left( {{{\bm e}_1} \cdot {{\hat {\bm k}}_2}} \right){{\hat {\bm k}}_2} \times {{\hat {\bm k}}_1}} \right],\\
{\left( {{{\dot {\bm e}}_2}} \right)_{\rm quad}} =&  - \frac{{{\gamma _2}}}{{\left\| {{{\bm G}_2}} \right\|}}\left[ {\left( {1 - e_1^2} \right)\left( {{{\hat {\bm k}}_1} \cdot {{\hat {\bm k}}_2}} \right){{\hat {\bm k}}_1} \times {{\bm e}_2} - 5\left( {{{\bm e}_1} \cdot {{\hat {\bm k}}_2}} \right){{\bm e}_1} \times {{\bm e}_2}} \right.\\
&\left.{+ \frac{1}{2}\left( {1 - 6e_1^2 - 5\left( {1 - e_1^2} \right){{\left( {{{\hat {\bm k}}_1} \cdot {{\hat {\bm k}}_2}} \right)}^2} + 25{{\left( {{{\bm e}_1} \cdot {{\hat {\bm k}}_2}} \right)}^2}} \right){{\hat {\bm k}}_2} \times {{\bm e}_2}} \right],
\end{aligned}
\end{equation*}
and the associated octupole-order contribution can be written as
\begin{equation*}
\begin{aligned}
{\left( {{{\dot {\bm G}}_1}} \right)_{\rm oct}} =& {\gamma _3}\left[ {\left( {{\cal B}{{\bm e}_2} + {\cal C}{{\hat {\bm k}}_2}} \right) \times {{\bm e}_1} + \left( {{\cal D}{{\bm e}_2} + {\cal E}{{\hat {\bm k}}_2}} \right) \times {{\hat {\bm k}}_1}} \right],\quad {\left( {{{\dot {\bm G}}_2}} \right)_{\rm oct}} =  - {\left( {{{\dot {\bm G}}_1}} \right)_{\rm oct}},\\
{\left( {{{\dot {\bm e}}_1}} \right)_{\rm oct}} =& \frac{{{\gamma _3}}}{{\left\| {{{\bm G}_1}} \right\|}}\left[ {\left( {1 - e_1^2} \right)\left( {{\cal A}{{\bm e}_1} + {\cal B}{{\bm e}_2} + {\cal C}{{\hat {\bm k}}_2}} \right) \times {{\hat {\bm k}}_1} + \left( {{\cal D}{{\bm e}_2} + {\cal E}{{\hat {\bm k}}_2}} \right) \times {{\bm e}_1}} \right],\\
{\left( {{{\dot {\bm e}}_2}} \right)_{\rm oct}} =& \frac{{{\gamma _3}}}{{\left\| {{{\bm G}_2}} \right\|}}\left[ {\left( {{\cal F} + {\cal C}{{\bm e}_1} \cdot {{\hat {\bm k}}_2} + {\cal E}\left( {{{\hat {\bm k}}_1} \cdot {{\hat {\bm k}}_2}} \right)} \right){{\bm e}_2} \times {{\hat {\bm k}}_2} + \left( {1 - e_2^2} \right)\left( {{\cal B}{{\bm e}_1} + {\cal D}{{\hat {\bm k}}_1}} \right) \times {{\hat {\bm k}}_2} + \left( {{\cal C}{{\bm e}_1} + {\cal E}{{\hat {\bm k}}_1}} \right) \times {{\bm e}_2}} \right],
\end{aligned}
\end{equation*}
where
\begin{equation*}
{\gamma _2} = \frac{{3{\cal G}{m_2}{\beta _1}a_1^2}}{{4 a_2^3{{\left( {1 - e_2^2} \right)}^{3/2}}}},\quad {\gamma _3} = \frac{{15}}{{64}}\frac{{{\cal G}{m_2}{\beta _1}a_1^3}}{{a_2^4{{\left( {1 - e_2^2} \right)}^{5/2}}}}\frac{{{m_0} - {m_1}}}{{{m_0} + {m_1}}},
\end{equation*}
and
\begin{equation*}
\begin{aligned}
{\cal A} &= 16\left( {{{\bm e}_1} \cdot {{\bm e}_2}} \right),\quad {\cal B} =  - \left[ {1 - 5\left( {1 - e_1^2} \right){{\left( {{{\hat {\bm k}}_1} \cdot {{\hat {\bm k}}_2}} \right)}^2} + 35{{\left( {{{\bm e}_1} \cdot {{\hat {\bm k}}_2}} \right)}^2} - 8e_1^2} \right],\\
{\cal C} &= 10\left( {1 - e_1^2} \right)\left( {{{\hat {\bm k}}_1} \cdot {{\bm e}_2}} \right)\left( {{{\hat {\bm k}}_1} \cdot {{\hat {\bm k}}_2}} \right) - 70\left( {{{\bm e}_1} \cdot {{\hat {\bm k}}_2}} \right)\left( {{{\bm e}_1} \cdot {{\bm e}_2}} \right),\quad {\cal D} = 10\left( {1 - e_1^2} \right)\left( {{{\bm e}_1} \cdot {{\hat {\bm k}}_2}} \right)\left( {{{\hat {\bm k}}_1} \cdot {{\hat {\bm k}}_2}} \right),\\
{\cal E} &= 10\left( {1 - e_1^2} \right)\left[ {\left( {{{\bm e}_1} \cdot {{\bm e}_2}} \right)\left( {{{\hat {\bm k}}_1} \cdot {{\hat {\bm k}}_2}} \right) + \left( {{{\bm e}_1} \cdot {{\hat {\bm k}}_2}} \right)\left( {{{\hat {\bm k}}_1} \cdot {{\bm e}_2}} \right)} \right],\quad {\cal F} = 5\left[ {{\cal B}\left( {{{\bm e}_1} \cdot {{\bm e}_2}} \right) + {\cal D}\left( {{{\hat {\bm k}}_1} \cdot {{\bm e}_2}} \right)} \right].
\end{aligned}
\end{equation*}

The dominant correction of general relativistic effects to the evolution of inner orbit is given by
\begin{equation*}
{\left( {{{\dot {\bm e}}_1}} \right)_{\rm gr}} = \frac{{3{\mu _1}{n_1}}}{{{c^2}{a_1}\left( {1 - e_1^2} \right)}}\left( {{{\hat {\bm k}}_1} \times {{\bm e}_1}} \right),
\end{equation*}
where $c$ is the speed of light and $n_1 = \sqrt{\mu_1/a_1^3}$ is the mean motion of the inner binary.

As all bodies are considered as oblate ellipsoids, the contribution from their rotational deformation (quadrupole-order approximation) can be written as
\begin{equation*}
\begin{aligned}
{\left( {{{\dot {\bm L}}_0}} \right)_{\rm rot}} &=  - {\alpha _0}\left( {{{\hat {\bm s}}_0} \cdot {{\hat {\bm k}}_1}} \right){{\hat {\bm k}}_1} \times {{\hat {\bm s}}_0},\quad {\left( {{{\dot {\bm L}}_1}} \right)_{\rm rot}} =  - {\alpha _1}\left( {{{\hat {\bm s}}_1} \cdot {{\hat {\bm k}}_1}} \right){{\hat {\bm k}}_1} \times {{\hat {\bm s}}_1},\quad {\left( {{{\dot {\bm L}}_2}} \right)_{\rm rot}} =  - {\alpha _2}\left( {{{\hat {\bm s}}_2} \cdot {{\hat {\bm k}}_2}} \right){{\hat {\bm k}}_2} \times {{\hat {\bm s}}_2},\\
{\left( {{{\dot {\bm G}}_1}} \right)_{\rm rot}} &=  - {\alpha _0}\left( {{{\hat {\bm s}}_0} \cdot {{\hat {\bm k}}_1}} \right){{\hat {\bm s}}_0} \times {{\hat {\bm k}}_1} - {\alpha _1}\left( {{{\hat {\bm s}}_1} \cdot {{\hat {\bm k}}_1}} \right){{\hat {\bm s}}_1} \times {{\hat {\bm k}}_1},\quad {\left( {{{\dot {\bm G}}_2}} \right)_{\rm rot}} =  - {\alpha _2}\left( {{{\hat {\bm s}}_2} \cdot {{\hat {\bm k}}_2}} \right){{\hat {\bm s}}_2} \times {{\hat {\bm k}}_2},\\
{\left( {{{\dot {\bm e}}_1}} \right)_{\rm rot}} &=  - \sum\limits_{i = 0,1} {\frac{{{\alpha _i}}}{{\left\| {{{\bm G}_1}} \right\|}}\left[ {\left( {{{\hat {\bm s}}_i} \cdot {{\hat {\bm k}}_1}} \right){{\hat {\bm s}}_i} \times {{\bm e}_1} + \frac{1}{2}\left( {1 - 5{{\left( {{{\hat {\bm s}}_i} \cdot {{\hat {\bm k}}_1}} \right)}^2}} \right){{\hat {\bm k}}_1} \times {{\bm e}_1}} \right]},\\
{\left( {{{\dot {\bm e}}_2}} \right)_{\rm rot}} &=  - \frac{{{\alpha _2}}}{{\left\| {{{\bm G}_2}} \right\|}}\left[ {\left( {{{\hat {\bm s}}_2} \cdot {{\hat {\bm k}}_2}} \right){{\hat {\bm s}}_2} \times {{\bm e}_2} + \frac{1}{2}\left( {1 - 5{{\left( {{{\hat {\bm s}}_2} \cdot {{\hat {\bm k}}_2}} \right)}^2}} \right){{\hat {\bm k}}_2} \times {{\bm e}_2}} \right],
\end{aligned}
\end{equation*}
where ${\alpha _{0,1,2}}$ are the structure-related coefficients, defined by
\begin{equation*}
{\alpha _0} = \frac{{3{\cal G}{m_0}{m_1}{J_{2,0}}R_0^2}}{{2a_1^3{{\left( {1 - e_1^2} \right)}^{3/2}}}},\quad {\alpha _1} = \frac{{3{\cal G}{m_0}{m_1}{J_{2,1}}R_1^2}}{{2a_1^3{{\left( {1 - e_1^2} \right)}^{3/2}}}},\quad {\alpha _2} = \frac{{3{\cal G}{(m_0+m_1)}{m_2}{J_{2,2}}R_2^2}}{{2a_2^3{{\left( {1 - e_2^2} \right)}^{3/2}}}},
\end{equation*}
with $R_i$ as the radius of each object.

Tidal effects include tidal deformation and tidal friction, whose contributions to the spin-orbit evolutions can be written as
\begin{equation*}
\begin{aligned}
{\left( {{{\dot {\bm L}}_0}} \right)_{\rm tide}} =& {K_0}{n_1}\left[ {{f_2}\left( {{e_1}} \right){{\hat {\bm k}}_1} + \left( {\frac{{{\Omega _{{\rm rot},0}}}}{{{n_1}}}} \right)\left( {\frac{1}{2}{\eta _1}{f_4}\left( {{e_1}} \right)\left( {{{\hat {\bm s}}_0} - \left( {{{\hat {\bm s}}_0} \cdot {{\hat {\bm k}}_1}} \right){{\hat {\bm k}}_1}} \right) - {f_1}\left( {{e_1}} \right){{\hat {\bm s}}_0} + \frac{{\left( {{{\bm e}_1} \cdot {{\hat {\bm s}}_0}} \right)\left( {6 + e_1^2} \right)}}{{4{{\left( {1 - e_1^2} \right)}^{9/2}}}}{{\bm e}_1}} \right)} \right],\\
{\left( {{{\dot {\bm L}}_1}} \right)_{\rm tide}} =& {K_1}{n_1}\left[ {{f_2}\left( {{e_1}} \right){{\hat {\bm k}}_1} + \left( {\frac{{{\Omega _{{\rm rot},1}}}}{{{n_1}}}} \right)\left( {\frac{1}{2}{\eta _1}{f_4}\left( {{e_1}} \right)\left( {{{\hat {\bm s}}_1} - \left( {{{\hat {\bm s}}_1} \cdot {{\hat {\bm k}}_1}} \right){{\hat {\bm k}}_1}} \right) - {f_1}\left( {{e_1}} \right){{\hat {\bm s}}_1} + \frac{{\left( {{{\bm e}_1} \cdot {{\hat {\bm s}}_1}} \right)\left( {6 + e_1^2} \right)}}{{4{{\left( {1 - e_1^2} \right)}^{9/2}}}}{{\bm e}_1}} \right)} \right],\\
{\left( {{{\dot {\bm L}}_2}} \right)_{\rm tide}} =& {K_2}{n_2}\left[ {{f_2}\left( {{e_2}} \right){{\hat {\bm k}}_2} + \left( {\frac{{{\Omega _{{\rm rot},2}}}}{{{n_2}}}} \right)\left( {\frac{1}{2}{\eta _2}{f_4}\left( {{e_2}} \right)\left( {{{\hat {\bm s}}_2} - \left( {{{\hat {\bm s}}_2} \cdot {{\hat {\bm k}}_2}} \right){{\hat {\bm k}}_2}} \right) - {f_1}\left( {{e_2}} \right){{\hat {\bm s}}_2} + \frac{{\left( {{{\bm e}_2} \cdot {{\hat {\bm s}}_2}} \right)\left( {6 + e_2^2} \right)}}{{4{{\left( {1 - e_2^2} \right)}^{9/2}}}}{{\bm e}_2}} \right)} \right],\\
{\left( {{{\dot {\bm G}}_1}} \right)_{\rm tide}} =&  - {\left( {{{\dot {\bm L}}_0}} \right)_{\rm tide}} - {\left( {{{\dot {\bm L}}_1}} \right)_{\rm tide}},\quad {\left( {{{\dot {\bm G}}_2}} \right)_{\rm tide}} =  - {\left( {{{\dot {\bm L}}_2}} \right)_{\rm tide}},\\
{\left( {{{\dot {\bm e}}_1}} \right)_{\rm tide}} =& \sum\limits_{i = 0,1} {\frac{{15}}{2}{n_1}{k_{2,i}}\left( {\frac{{{m_{1 - i}}}}{{{m_i}}}} \right){{\left( {\frac{{{R_i}}}{{{a_1}}}} \right)}^5}{f_4}\left( {{e_1}} \right){{\hat {\bm k}}_1} \times {{\bm e}_1}}\\
&- \sum\limits_{i = 0,1} {\frac{{{K_i}}}{{{\beta _1}a_1^2}}\left[ {{f_4}\left( {{e_1}} \right)\frac{{{\Omega _{{\rm rot},i}}}}{{2{n_1}}}\left( {{{\bm e}_1} \cdot {{\hat {\bm s}}_i}} \right){{\hat {\bm k}}_1} - \left( {\frac{{11}}{2}{f_4}\left( {{e_1}} \right)\left( {{{\hat {\bm s}}_i} \cdot {{\hat {\bm k}}_1}} \right)\frac{{{\Omega _{{\rm rot},i}}}}{{{n_1}}} - 9{f_5}\left( {{e_1}} \right)} \right){{\bm e}_1}} \right]},\\
{\left( {{{\dot {\bm e}}_2}} \right)_{\rm tide}} =& \frac{{15}}{2}{n_2}{k_{2,2}}\left( {\frac{{{m_0+m_1}}}{{{m_2}}}} \right){\left( {\frac{{{R_2}}}{{{a_2}}}} \right)^5}{f_4}\left( {{e_2}} \right){{\hat {\bm k}}_2} \times {{\bm e}_2}\\
&- \frac{{{K_2}}}{{{\beta _2}a_2^2}}\left[ {{f_4}\left( {{e_2}} \right)\frac{{{\Omega _{{\rm rot},2}}}}{{2{n_2}}}\left( {{{\bm e}_2} \cdot {{\hat {\bm s}}_2}} \right){{\hat {\bm k}}_2} - \left( {\frac{{11}}{2}{f_4}\left( {{e_2}} \right)\left( {{{\hat {\bm s}}_2} \cdot {{\hat {\bm k}}_2}} \right)\frac{{{\Omega _{{\rm rot},2}}}}{{{n_2}}} - 9{f_5}\left( {{e_2}} \right)} \right){{\bm e}_2}} \right],
\end{aligned}
\end{equation*}
where $\eta_1 = \sqrt{1-e_1^2}$, $\eta_2 = \sqrt{1-e_2^2}$, and $n_2 = \sqrt{\mu_2/a_2^3}$. It is noted that the terms dependent on $K_{i}$ correspond to dissipative effects of tidal friction, and the terms independent on $K_{i}$ corresponds to conservative effects of tide deformation. The time delay-related coefficients are defined by
\begin{equation*}
{K_0} = {3{k_{2,0}}}{\Delta {t_0}}\frac{{{\cal G}m_1^2R_0^5}}{{a_1^6}},\quad {K_1} = {3{k_{2,1}}}{\Delta {t_1}}\frac{{{\cal G}m_0^2R_1^5}}{{a_1^6}},\quad {K_2} = {3{k_{2,2}}}{\Delta {t_2}}\frac{{{\cal G}(m_0+m_1)^2R_2^5}}{{a_2^6}},
\end{equation*}
with $\Delta t_{0,1,2}$ as the time delay, and the eccentricity-related coefficients are given by
\begin{equation*}
\begin{aligned}
{f_1}\left( e \right) &= \frac{{1 + 3{e^2} + \frac{3}{8}{e^4}}}{{{{\left( {1 - {e^2}} \right)}^{9/2}}}},\quad {f_2}\left( e \right) = \frac{{1 + \frac{{15}}{2}{e^2} + \frac{{45}}{8}{e^4} + \frac{5}{{16}}{e^6}}}{{{{\left( {1 - {e^2}} \right)}^6}}},\quad {f_3}\left( e \right) = \frac{{1 + \frac{{31}}{2}{e^2} + \frac{{255}}{8}{e^4} + \frac{{185}}{{16}}{e^6} + \frac{{25}}{{64}}{e^8}}}{{{{\left( {1 - {e^2}} \right)}^{15/2}}}},\\
\quad {f_4}\left( e \right) &= \frac{{1 + \frac{3}{2}{e^2} + \frac{1}{8}{e^4}}}{{{{\left( {1 - {e^2}} \right)}^5}}},\quad {f_5}\left( e \right) = \frac{{1 + \frac{{15}}{4}{e^2} + \frac{{15}}{8}{e^4} + \frac{5}{{64}}{e^6}}}{{{{\left( {1 - {e^2}} \right)}^{13/2}}}}.
\end{aligned}
\end{equation*}

\section{Some useful expressions}
\label{Appendix_B}

In terms of orbital elements, the unitary vector of orbital angular momentum and eccentricity vector can be expressed as
\begin{equation}\label{Eq_A2}
{\hat{\bm k}_1}= \left( {\begin{array}{*{20}{c}}
{\sin {i_1}\sin {\Omega _1}}\\
{ - \sin {i_1}\cos {\Omega _1}}\\
{\cos {i_1}}
\end{array}} \right),\quad {\hat{\bm e}_1} = \left( {\begin{array}{*{20}{c}}
{\cos {\Omega _1}\cos {\omega _1} - \sin {\Omega _1}\sin {\omega _1}\cos {i_1}}\\
{\sin {\Omega _1}\cos {\omega _1} + \cos {\Omega _1}\sin {\omega _1}\cos {i_1}}\\
{\sin {\omega _1}\sin {i_1}}
\end{array}} \right),\quad {\hat {\bm q}_1} = {\hat {\bm k}_1} \times {\hat {\bm e}_1}.
\end{equation}
Similar expressions can be obtained for ${\hat{\bm k}_2}$, ${\hat{\bm e}_2}$ and ${\hat{\bm q}_2}$ in terms of $i_2$, $\Omega_2$ and $\omega_2$.

In an inertial coordinate system (the invariant plane coordinate system is adopted in this work), the spin axes $\hat {\bm s}_{0,1,2}$ can be characterized by orientation angles as follows:
\begin{equation*}
{\hat {\bm s}_j} = \left( {\begin{array}{*{20}{c}}
{\sin {K_j}\sin {\varphi _j}}\\
{ - \sin {K_j}\cos {\varphi _j}}\\
{\cos {K_j}}
\end{array}} \right),\quad j=0,1,2.
\end{equation*}
Based on the expressions of ${\hat {\bm s}_{0,1,2}}$ and ${\hat{\bm k}_{1,2}}$, we could get the transformation between $(\theta_{j},\phi_{j})$ and $(K_{j},\varphi_{j})$ for the oblate bodies $m_0$ and $m_1$ as follows:
\begin{equation*}
\left\{ \begin{array}{l}
\sin {\theta _j}\sin {\phi _j} = \sin {K_j}\sin \left( {{\varphi _j} - {\Omega _1}} \right)\\
\sin {\theta _j}\cos {\phi _j} = \sin {K_j}\cos {i_1}\cos \left( {{\varphi _j} - {\Omega _1}} \right) - \cos {K_j}\sin {i_1}\\
\cos {\theta _j} = \cos {K_j}\cos {i_1} + \sin {K_j}\sin {i_1}\cos \left( {{\varphi _j} - {\Omega _1}} \right)
\end{array} \right.,\;\left\{ \begin{array}{l}
\sin {K_j}\sin \left( {{\varphi _j} - {\Omega _1}} \right) = \sin {\theta _j}\sin {\phi _j}\\
\sin {K_j}\cos \left( {{\varphi _j} - {\Omega _1}} \right) = \cos {\theta _j}\sin {i_1} + \sin {\theta _j}\cos {i_1}\cos {\phi _j}\\
\cos {K_j} = \cos {\theta _j}\cos {i_1} - \sin {\theta _j}\sin {i_1}\cos {\phi _j}
\end{array} \right.
\end{equation*}
with $j=0,1$. Similarly, the transformation between $(\varepsilon_{2},\phi_{2})$ and $(K_{2},\varphi_{2})$ for the planet $m_2$ can be written as
\begin{equation*}
\left\{ \begin{array}{l}
\sin {\varepsilon _2}\sin {\phi _2} = \sin {K_2}\sin \left( {{\varphi _2} - {\Omega _2}} \right)\\
\sin {\varepsilon _2}\cos {\phi _2} = \sin {K_2}\cos {i_2}\cos \left( {{\varphi _2} - {\Omega _2}} \right) - \cos {K_2}\sin {i_2}\\
\cos {\varepsilon _2} = \cos {K_2}\cos {i_2} + \sin {K_2}\sin {i_2}\cos \left( {{\varphi _2} - {\Omega _2}} \right)
\end{array} \right.,\;\left\{ \begin{array}{l}
\sin {K_2}\sin \left( {{\varphi _2} - {\Omega _2}} \right) = \sin {\varepsilon _2}\sin {\phi _2}\\
\sin {K_2}\cos \left( {{\varphi _2} - {\Omega _2}} \right) = \cos {\varepsilon _2}\sin {i_2} + \sin {\varepsilon _2}\cos {i_2}\cos {\phi _2}\\
\cos {K_2} = \cos {\varepsilon _2}\cos {i_2} - \sin {\varepsilon _2}\sin {i_2}\cos {\phi _2}
\end{array} \right.
\end{equation*}


\bibliography{mybib}{}
\bibliographystyle{aasjournal}



\end{document}